\documentclass[prl,aps,showpacs,twocolumn,unsortedaddress,longbibliography,superscriptaddress]{revtex4-2}
\usepackage{graphics, bm, xspace}
\usepackage{graphicx}
\usepackage{psfrag}
\usepackage{amsmath}
\usepackage{amssymb}
\usepackage{epsfig}
\usepackage{subfigure}
\usepackage{grffile}
\usepackage{times}
\usepackage{color}
\usepackage{multirow}
\usepackage[bookmarks=false,linkcolor=blue,urlcolor=blue,colorlinks,citecolor=blue]{hyperref}
\usepackage{float}
\usepackage{gensymb}
\usepackage{mhchem}

\newcommand{\beq}    {\begin{equation}}
\newcommand{\enq}    {\end{equation}}
\newcommand{\ceq}[1] {(\ref{#1})}

\newcommand{\bb}     {{\bf b}}

\newcommand{\kk}     {{\bf k}}

\newcommand{\qq}     {{\bf q}}

\newcommand{\KK}     {{\bf K}}

\newcommand{\rr}     {{\bf r}}

\newcommand{\ssigma}     {{\boldsymbol\sigma}}

\newcommand{\rs}{$\rho_s$\xspace}
\newcommand{\rhos}{$\rho_s$\xspace}

\begin{document}

\title{Quantum-metric-enabled exciton condensate in double twisted bilayer graphene}

\author{Xiang Hu}
\affiliation{Department of Physics, William \& Mary, Williamsburg, VA 23187, USA}

\author{Timo Hyart}
\affiliation{International Research Centre MagTop, Institute of Physics, Polish Academy of Sciences, Aleja Lotnikow 32/46, PL-02668 Warsaw, Poland}
\affiliation{Department of Applied Physics, Aalto University, 00076 Aalto, Espoo, Finland}

\author{Dmitry I. Pikulin}
\affiliation{Microsoft Quantum, Redmond, Washington 98052, USA}
\affiliation{Microsoft Quantum, Station Q, Santa Barbara, California
	93106-6105, USA}

\author{Enrico Rossi}
\affiliation{Department of Physics, William \& Mary, Williamsburg, VA 23187, USA}

\date{\today}


\begin{abstract}
Flat-band systems are a promising platform for realizing exotic collective ground states with spontaneously broken symmetry because the electron-electron interactions dominate over the kinetic energy.  A collective ground state of particular interest is the chased after exciton condensate (EC).
However, in flat band systems other collective ground states can compete with an EC phase, and the conventional treatment of the effect of thermal
and quantum fluctuations predicts the EC phase should be unstable.
Here, using double twisted bilayer graphene (TBLG) heterostructures as an example, we show that for realistic interaction strengths
the EC phase is favored with respect to other TBLG's phases -- orbital magnetism and superconductivity-- when the TBLGs have opposite doping,
and that the quantum metric of the Bloch wave functions stabilizes the EC, reversing the conclusion that would be drawn from the conventional approach
in which quantum metric contributions are neglected.
Our results suggest that the quantum metric plays a critical role in determining the
stability of exciton condensates in double layers formed by systems with flat-bands.
\end{abstract}

\maketitle

An exciton is a bosonic quasiparticle formed by an electron (e) bound to a hole (h).
A large number of excitons can become phase coherent and form a collective state
known as exciton condensate (EC)~\cite{Keldysh1965,Halperin1968}.
Already in the mid 70's it was proposed~\cite{lozovik1975,lozovik1976}
that spatially separating electrons and holes should facilitate the formation of a thermodynamically stable EC.
Such separation can be realized
in e-h semiconductor double layers, in which
a thin dielectric separates the layers
and distinct metal gates are used
to create an excess density of electrons in one layer which equals the excess density of holes in the other one.
Great advances in the fabrication of heterostructures made possible the realization of several novel double layers in which ECs could
be realized~\cite{Min2008,Joglekar2008,kharitonov2008,Zhang2013,Perali2013, Neilson14, Pikulin14, Fogler2014a, Pikulin16, li2017,Debnath2017,su2017, Du17,Burg2018,Tutuc2019,Wang2019,Liu2019, Kwan2020,Kwan2020a,Shimazaki2020}.
It was proposed that ECs could be formed in graphene double layers~\cite{Min2008,Joglekar2008},
but experimentally no strong signatures have been observed, so far.
It was then proposed that ECs could be realized in systems based on double bilayer graphene (BLG)~\cite{Zhang2013,Perali2013,su2017}, given
that at low energies BLG's bands are qualitatively flatter than graphene's, and recent experiments
show signatures that are consistent with
the formation of an EC~\cite{Burg2018}.
These results, combined with the ones for quantum Hall (QH) bilayers~\cite{spielman2000,stern2001,eisenstein2004,rossi2005,finck2011,hyart2011},
in which the kinetic term of each layer is completely quenched,
would suggest that, in general, the formation of an EC is favored in bilayers formed by 2D systems
with flat bands.
As a consequence, double twisted bilayer graphene (TBLG), in which the
bands can be made extremely flat by tuning the twist angle $\theta$ between graphene
sheets~\cite{dossantos2007,morell2010,Bistritzer2011a,Cao2018,Cao2018b,Yankowitz2019,Lu2019,Andrei2020}
appears to be an ideal system to seek the realization of ECs without external magnetic fields.
This expectation, however, is in part naive.
First, the flatness of the bands is associated with strong screening
of the interlayer Coulomb interaction that is the driver of the EC instability.
This obstacle can be overcome by tuning the system into the
strong coupling regime, where the e-(h-)densities are sufficiently small so that the coherence length $\xi$ of the EC is smaller
than the average distance between particles \cite{Neilson14}.
Second, the stiffness ($\rho_s$) of the EC, i.e. its robustness against thermal and quantum fluctuations, is conventionally expected
to decrease as the bands become flatter and ultimately vanish in the limit of perfectly flat bands.

\begin{figure}[htb]
\centering
\includegraphics[width=\linewidth]{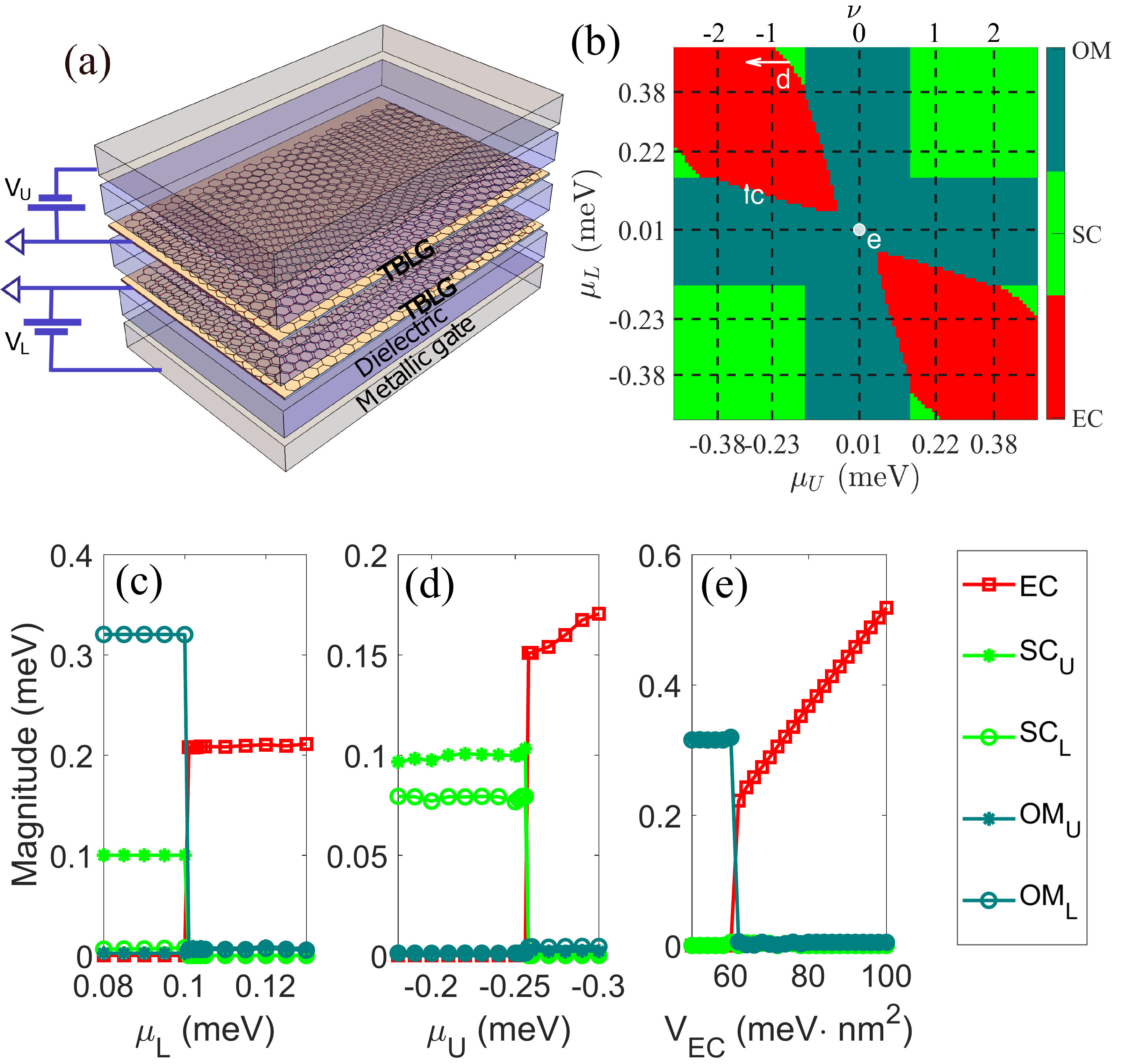}
\caption{
 	     (a)   Proposed experimental setup.
	     (b)   Phase diagram of double-TBLG as a function of $\mu_U$ and $\mu_L$ for $\theta=1.00\degree$.
         (c,d) Phase transitions as a function of dopings along the arrows shown in (b).
         (e)   Phase transition as a function of $V_{\rm EC}$ at $\nu_U=\nu_L=0$. The legend ${\rm SC}({\rm OM})_{{\rm U}({\rm L})}$ represents the SC~(OM) phase in the Upper~(Lower) TBLG.
         }
\label{fig:sketch}
\end{figure}

In this work we show that the second obstacle in general might not be present if one considers the contribution
to \rhos due to the quantum metric of the eigenstates of the EC.
We consider the specific case of  double layers formed by an
e-doped TBLG and a h-doped TBLG separated by a thin insulating barrier  [Fig.~\ref{fig:sketch}(a)].
We first perform a mean field calculation, in which
the order parameters for the EC, superconductivity (SC)  and orbital magnetism (OM)
are treated on equal footing, to identify the regions of the
the phase diagram  as a function of dopings in the upper (U) and lower (L) TBLG  where the EC is favored.
We then calculate $\rho_s$ for the EC and show that the
contribution to it due to the quantum metric is essential to make it positive and
therefore to stabilize the EC.
In addition, we describe how $\rho_s$ depends on the twist angle and find that
the most favorable twist angle $\theta$ to realize a stable EC is not the magic angle.
We also obtain the Berezinskii-Kosterlitz-Thouless (BKT) temperature $T_{\mathrm{BKT}}$~\cite{Berezinski1971,Kosterlitz1973}
as function of $\theta$.
Considering that most systems with almost flat bands are multiband systems, our results
have universal relevance for the understanding of the conditions necessary to realize ECs:
they show that to realize an EC
in 2D bilayers
the flatness of the bands of the layers must
be accompanied by a significant quantum metric contribution to the EC's stiffness. Our results also allow to understand
in a new light the conditions that make possible the realization and observation of ECs in QH  bilayers~\cite{Yang1994,Moon1995}.

The double TBLG system is described by the Hamiltonian
$
 \hat H = \hat H^U + \hat H^L + \hat H_{\rm int}
$
where $\hat H^{U/L}$ is the single-particle Hamiltonian for the U/L TBLG and $H_{\rm int}$ describes the e-e interactions.
We assume $\theta$ to be the same for the two TBLGs.
For small $\theta$ the low energy states of a TBLG are well described
by an effective tight-binding Hamiltonian {\em in momentum space}
with the lattice sites $\{\bb=m_1\bb_1+m_2\bb_2\}$ corresponding to the reciprocal lattice vectors of the moir\'e lattice. The on-site Hamiltonians describe the Dirac points of graphene with Fermi velocity $v_F=10^6$ m/s,
and the nearest-neighbor hopping matrices $T_i$
describe the coupling between the layers with tunneling strength $w=118$~meV \cite{Bistritzer2011a,SM, Jung2014,Carr2018}.
Here $\bb_1=(\sqrt{3}Q, 0), \bb_2 =(\sqrt{3}Q/2, 3Q/2)$, ${m_1, m_2}\in\mathbb{Z}$, $Q=(8\pi/3a_0)\sin(\theta/2)$ and $a_0$
is the lattice constant of graphene.
Recent experimental and theoretical results suggest that for a single TBLG the strongest
instabilities are orbital-magnetism (OM), characterized
by a finite polarization in sublattice space, and superconductivity (SC) ~\cite{Xie2020, PeltonenPhysRevB.98.220504,Wu2018}.
We therefore  decouple
the interactions within the same TBLG
via the mean fields
$\Delta^{{\rm OM},{\rm SC}}_{\bb l\sigma l'\sigma'}(l=l',\sigma=\sigma')$,
where the indices $l, l'$ ($\sigma, \sigma'$) correspond to the layer (sublattice) degrees of freedom freedom within the U or L TBLG~\cite{SM}.
The interaction between electrons in different TBLGs
is decoupled via the EC mean field
$\Delta^{\rm EC}_{\bb l\sigma l'\sigma'}$.
We assume the EC, SM, and OM phases obey the spin-rotation symmetry.
Given the flatness of TBLG's low energy bands,
in the mean-field approximation all the interactions can be replaced by effective {\em local} interactions~\cite{SM}.
We denote the strengths of the effective local interaction in the OM, SC and EC channels as $V_{\rm OM}$, $V_{\rm SC}$ and $V_{\rm EC}$, respectively.
We expect $V_{\rm OM}>V_{\rm SC} \sim V_{\rm EC}$, but
it is challenging to estimate the precise values of the interaction strengths because of the interplay of
screening effects and collective instabilities. Thus, we adopt a pragmatic approach: we set
$V_{\rm OM}=130$~meV$\cdot$nm$^2$, and $V_{\rm SC}=75$~meV$\cdot$nm$^2$
so that the corresponding critical temperatures $T^{\rm OM}_c$ and $T^{\rm SC}_c$
are in  good agreement with the experimental observations~\cite{Cao2018,Lu2019},
and consider different range of values for $V_{\rm EC}$,  $60-100$~meV$\cdot$nm$^2$,
for which $T^{\rm EC}_c\sim 1-4$~K, and the system is in a strong coupling regime
where the screening does not prevent the formation of the EC.

The gap equations for each order parameter (OP) $\Delta^{\rm OP}_{\bar\alpha}$, where ${\rm OP}=\{\rm OM,SC,EC\}$, and $\bar\alpha$ is a collective index,
can be linearized close to the critical temperature $T^{\rm OP}_c$:
$\Delta^{\rm OP}_{\bar\alpha}=\sum_{\bar\beta} \chi_{{\bar\alpha}{\bar\beta}}^{\rm OP}\Delta^{\rm OP}_{\bar\beta}$,
where $\chi_{\bar{\alpha}\bar{\beta}}^{\rm OP}$ is
the bare susceptibility, independent of $\Delta^{\rm OP}_{\bar\alpha}$.
$T^{\rm OP}_c$ is obtained as the temperature $T$
for which the largest eigenvalue of
$\chi_{\bar{\alpha}\bar{\beta}}^{\rm OP}$
is equal to 1.
The expressions of $\chi_{\bar{\alpha}\bar{\beta}}^{\rm OP}$ for each phase are given in \cite{SM}.
In Fig.~\ref{fig:sketch}(b) we show the phase diagram, as function of doping in each TBLG, for $V_{\rm EC}=60$~meV$\cdot$nm$^2$,
obtained by identifying the highest $T^{\rm OP}_c$.
We have verified
for several $(\mu_U,\mu_L)$ value pairs
that the results obtained from the linearized and non-linearized gap equations are consistent.
Close to $\nu_U=\nu_L=0$ the correlated insulating phase OM is favored, whereas introducing equal electron densities in the two TBLGs $\mu_L\sim\mu_U$ favors the SC phase \cite{Black17}.
When the excess density of electrons in one TBLG equals the excess density of holes in the other TBLG, $\mu_U\sim-\mu_L$, the EC becomes dominant.
In our system the EC is formed by states in physically different TBLGs,
no pairing between states in bands with opposite Chern number is assumed,
and so the topology of the low energy bands does not
penalize the formation of a uniform {\em inter-TBLG} EC state~\cite{Bultinck2020a}.

To investigate the possible coexistence of ordered phases~\cite{Ojajarvi2018}
we have solved across several phase boundaries the full non-linear gap equations in which all the order parameters are allowed to be nonzero.
We used large numbers of random initial conditions and identified the solution with the smallest total energy as the ground state.
Fig.~\ref{fig:sketch}~(c) and (d)
show the evolution of the order parameters across the OM/EC and SC/EC phase boundaries, respectively.
In both cases the results suggest that the system undergoes a first-order quantum phase transition as the dopings are varied
in Fig.~\ref{fig:sketch}(b).
Fig.~\ref{fig:sketch}(e) shows the evolution of the order parameters as a function of $V_{\rm EC}$ at the neutrality point.
Also in this case the transition appears to be first order.
Figure~\ref{fig:sketch}(e) suggest that for $V_{\rm EC}>60$~mev$\cdot$nm$^2$ the EC is favored in a significant region of the $(\mu_U,\mu_L)$
plane.
In the reminder we focus on the
$\mu_L=-\mu_U\equiv\mu$ regime, with $\mu$ sufficiently large,
and set $V_{\rm EC}=100~$meV$\cdot$nm$^2$
so that, at the mean-field level, the EC phase is dominant.
To simplify the notation in the sections below
the ${\rm EC}$ label is implied.

\begin{figure}[!!!t]
 \begin{center}
  \centering
  \includegraphics[width=\columnwidth]{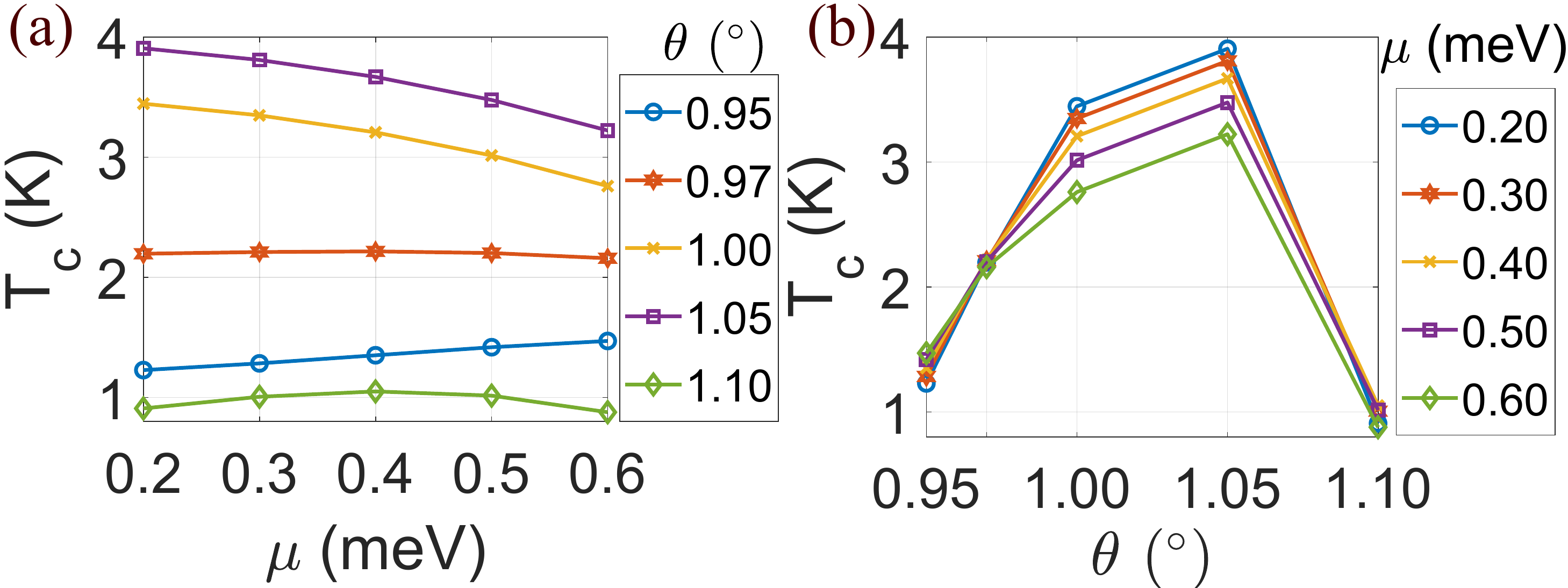}
  \caption{
           (a) $T_c$ as a function of $\mu=\mu_L=-\mu_U$ and different values of twist angle $\theta$.
           (b) $T_c$ as a function of $\theta$ and different values of $\mu$.
         }
  \label{fig:Tc}
 \end{center}
\end{figure}

\begin{figure}[!!!t]
 \begin{center}
  \centering
  \includegraphics[width=\columnwidth]{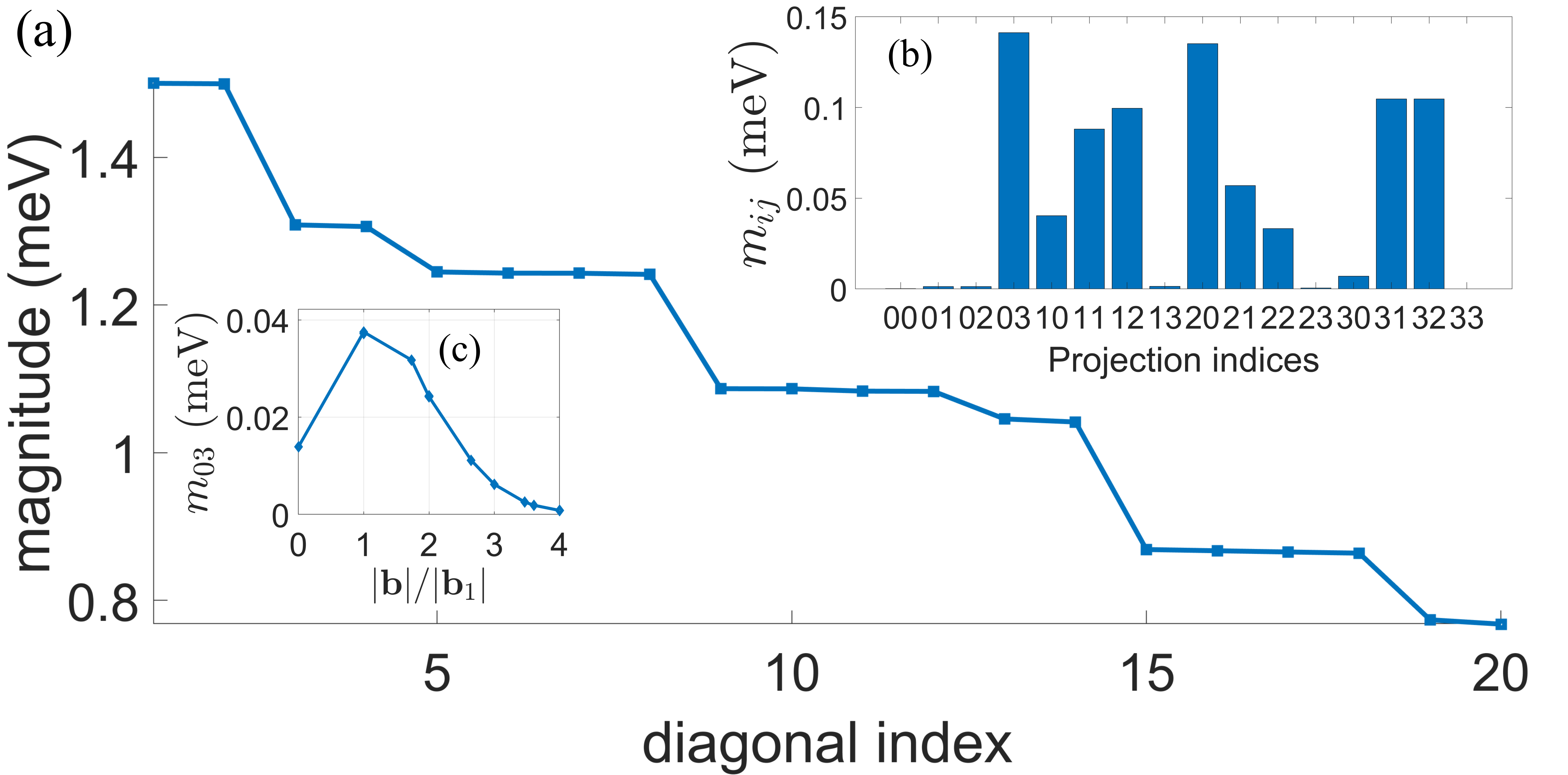}
  \caption{
           (a) The first twenty singular values of the SVD decomposition $\Delta_{\bb_l\sigma l'\sigma'} = USV^\dagger$.
           (b) Amplitudes of the order parameter components $m_{ij}$.
           (c) Scaling with $|\bb|$ of $m_{03}$. Here $\theta=1.05\degree$ and $\mu=0.30$ meV.
}
  \label{fig:delta}
 \end{center}
\end{figure}

Fig.~\ref{fig:Tc} shows how $T_c$ scales with $\mu$ and $\theta$ close to the magic angle $\theta_M=1.05 \degree$.
$T_c$ is largest when $\theta=\theta_M$, twist angle for which the bands are flattest,
and decreases quickly when $\theta$ is tuned away from $\theta_M$. %
The solution of the gap equation reveals that $\Delta_{\bb l\sigma l'\sigma'}$
has several non-zero components.
We performed the singular value decomposition (SVD), $\Delta_{\bb l\sigma l'\sigma'} = USV^\dagger$,
where $S$ is a diagonal matrix whose diagonal elements are the {\em singular values} of $\Delta_{\bb l\sigma l'\sigma'}$.
Fig.~\ref{fig:delta}(a) shows that the largest 20 singular values (in total we have 484 singular values~\cite{SM})
are of comparable size confirming the multi-component nature of the order parameter.

To better understand the orbital structure of $\Delta_{\bb l\sigma l'\sigma'}$
we calculated its projections on the 4$\times$4 matrices $\kappa_i\otimes\sigma_j$ as $m_{ij}=[\sum_{\bb}\|a^{(\bb)}_{ij}\|^2]^{1/2}$,
$a^{(\bb)}_{ij}=(1/4){\rm Tr}[\Delta_{\bb l\sigma l'\sigma'} \kappa_i\otimes\sigma_j$],
where $\kappa_i$ ($\sigma_i$) are the Pauli matrices in the layer (sublattice) space.
We see, Fig.~\ref{fig:delta}~(b), that
$m_{03}$ is the largest projection, but several other projections are comparable to it.
The fairly even distribution of the EC's order parameter over different orbital channels
is paralleled by its fairly slow decay with
$|\bb|$, see Fig.~\ref{fig:delta}(c).
These results
are consistent with the SVD's result that $\Delta_{\bb l\sigma l'\sigma'}$
describes a multi-component order parameter.
This is in contrast with the results for the case of
superconducting pairing in isolated TBLG where the pairing is dominated by a single channel and the magnitude of the order parameter decreases quickly with $|\bb|$~\cite{Wu2018,Hu2019}.

\begin{figure}[!!!t]
 \begin{center}
  \centering
  \includegraphics[width=\columnwidth]{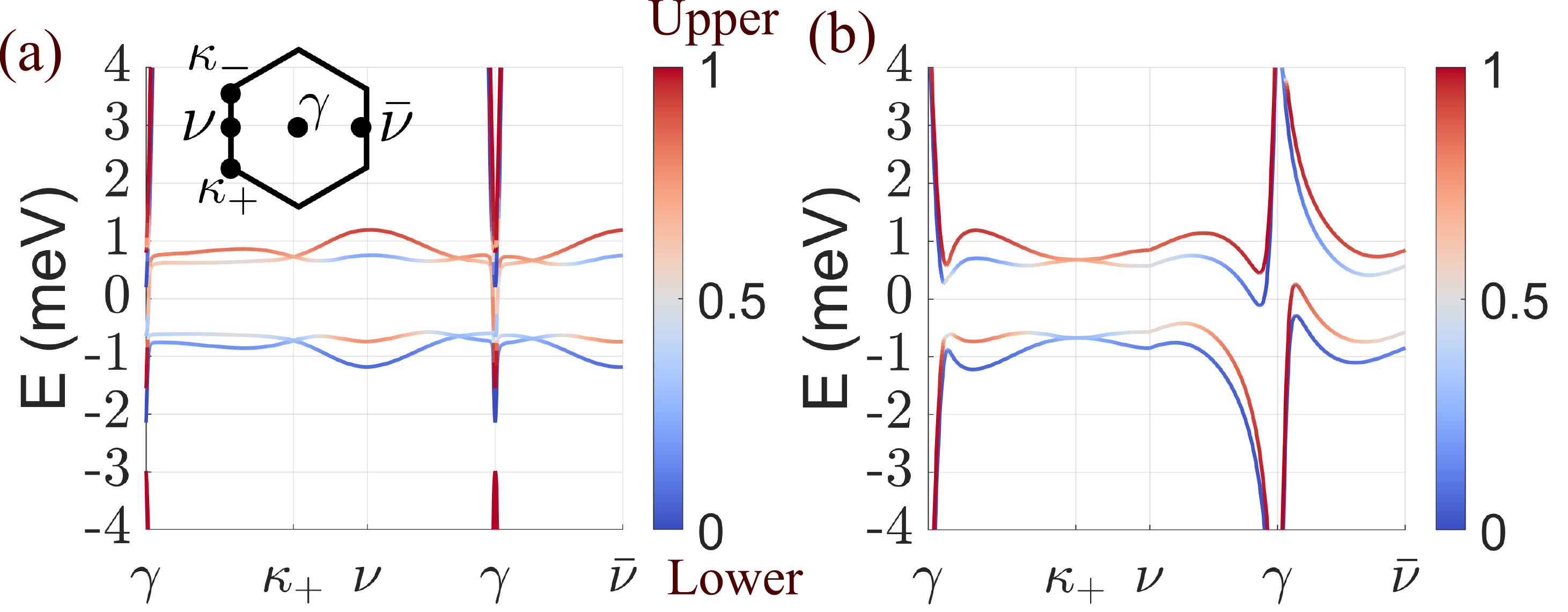}
  \caption{
           Band structures in the EC phase at $T=0$ and $\mu=0.30$ meV for (a) $\theta=1.05^\circ$ and (b) $\theta=1.00^\circ$. The colorbar indicates how much the eigenstate is localized in the $U$/$L$ TBLG. The inset in (a) shows the moir\'e Brillouin zone.}
  \label{fig:bands}
 \end{center}
\end{figure}

Fig.~\ref{fig:bands} shows the low energy bands along the $\gamma-\kappa_+ -\nu-\gamma-\bar\nu$ path in the moir\'e
Brillouin zone (BZ)~\cite{SM} for $\theta=1.05\degree$, and $\theta=1.00\degree$, in the presence of the EC condensate.
For $\theta=1.05\degree$ the very large Fermi velocity of the low energy bands at the $\gamma$
point prevents the EC from opening a gap at this point. As $\theta$ is tuned away from $\theta_M$
the singularity at the $\gamma$ point morphs into two very small e-h pockets, Fig.~\ref{fig:bands}(b).
The results of Fig.~\ref{fig:bands}(a,b) show that in double layer TBLG the EC is expected to be, strictly speaking, gapless.
However, given that the gapless nature is due to a very small number of states close to a single point of the BZ, the density of states is very negligible within the EC's gap (see \cite{SM}), and so we expect that the transition to the EC state could be clearly observed in transport and spectroscopy measurements.

We now consider the stability of the EC with respect to fluctuations.
The dominant fluctuations are the ones of the phase, $\varphi(\rr)$, of the order parameter:
$\Delta\to \Delta e^{i\varphi(\rr)}$.
Expanding the action in the long-wavelength limit around the saddle point identified by the mean-field solution we have
$S = \hat S_0 +\int d\tau \int d\rr\frac{1}{2}\rho^s_{\alpha\beta}\partial_{r_\alpha}\varphi\partial_{r_\beta}\varphi$,
where $S_0$ is the action at the saddle point, and
$\rho^s_{\alpha\beta}$ is the $\alpha\beta$ component of the EC's stiffness.
The EC is stable when $\rho^s_{\alpha\beta}$ is positive-definite.
For a multiband system $\rho^s_{\alpha\beta}$ is given by the general expression~\cite{Peotta2015,Liang2017}:
\begin{eqnarray}\label{eq:StiffTimo}
\rho^s_{\alpha\beta}&=&\sum_{{\bf k},i,j}\frac{n_F(E_j)-n_F(E_i)}{E_i-E_j}
\left(\frac{1}{4A}\langle\psi_i|\hat{v}_{\alpha}|\psi_j\rangle
\langle\psi_j|\hat{v}_{\beta}|\psi_i\rangle \right.
\nonumber\\
&&
-\left.\frac{1}{A}\langle\psi_i|\hat{v}_{cf,\alpha}|\psi_j\rangle
\langle\psi_j|\hat{v}_{cf,\beta}|\psi_i\rangle\right),
 \label{eq:rhos}
\end{eqnarray}
where
$E_i$ ($|\psi_i\rangle$) are the eigenvalues (eigenstates) of the mean-field Hamiltonian $H_\mathrm{MF}$, $n_F(E)$  is the Fermi-Dirac distribution, $A$ is the area of the sample,
$\hat{v}_{\alpha}({\bf k})=\partial H_\mathrm{MF}/\partial k_{\alpha}$ and
$\hat{v}_{cf,\alpha}({\bf k})=(1/2)\gamma_z\partial H_\mathrm{MF}/\partial k_{\alpha}$
are the components of the regular and counterflow velocity operators, respectively, $\gamma_z$ is the Pauli matrix acting in the U/L subspace, and $\kk=(k_x,k_y)$ is the Bloch wave vector. In our case, $\rho^s_{xy}=\rho^s_{yx}=0$, and $\rho^s_{xx}=\rho^s_{yy}\equiv\rho_s$.
For a multi-band system
we can identify a conventional contribution,  to \rs,  $\rho^{s,{\rm conv}}$, arising almost exclusively from intraband terms
({\it same band index in the electron or hole subspace}),
and a "geometric" contribution,
$\rho^{s,{\rm geom}}$,
due to interband terms
({\it different band indexes in both the electron and hole subspaces}), and write
$
 \rho^s = \rho^{s,{\rm conv}} + \rho^{s,{\rm geom}}.
$
Because
$\rho^{s,{\rm geom}}$ is closely connected to the quantum metric of the Hilbert space spanned by the eigenstates of
$H_\mathrm{MF}$~\cite{Peotta2015,Liang2017,Hu2019,Fang2020,Julku2020, cao2020quantum}, it is often called a geometric contribution to the superfluid stiffness.

\begin{figure}[!!!t]
 \begin{center}
  \centering
  \includegraphics[width=\columnwidth]{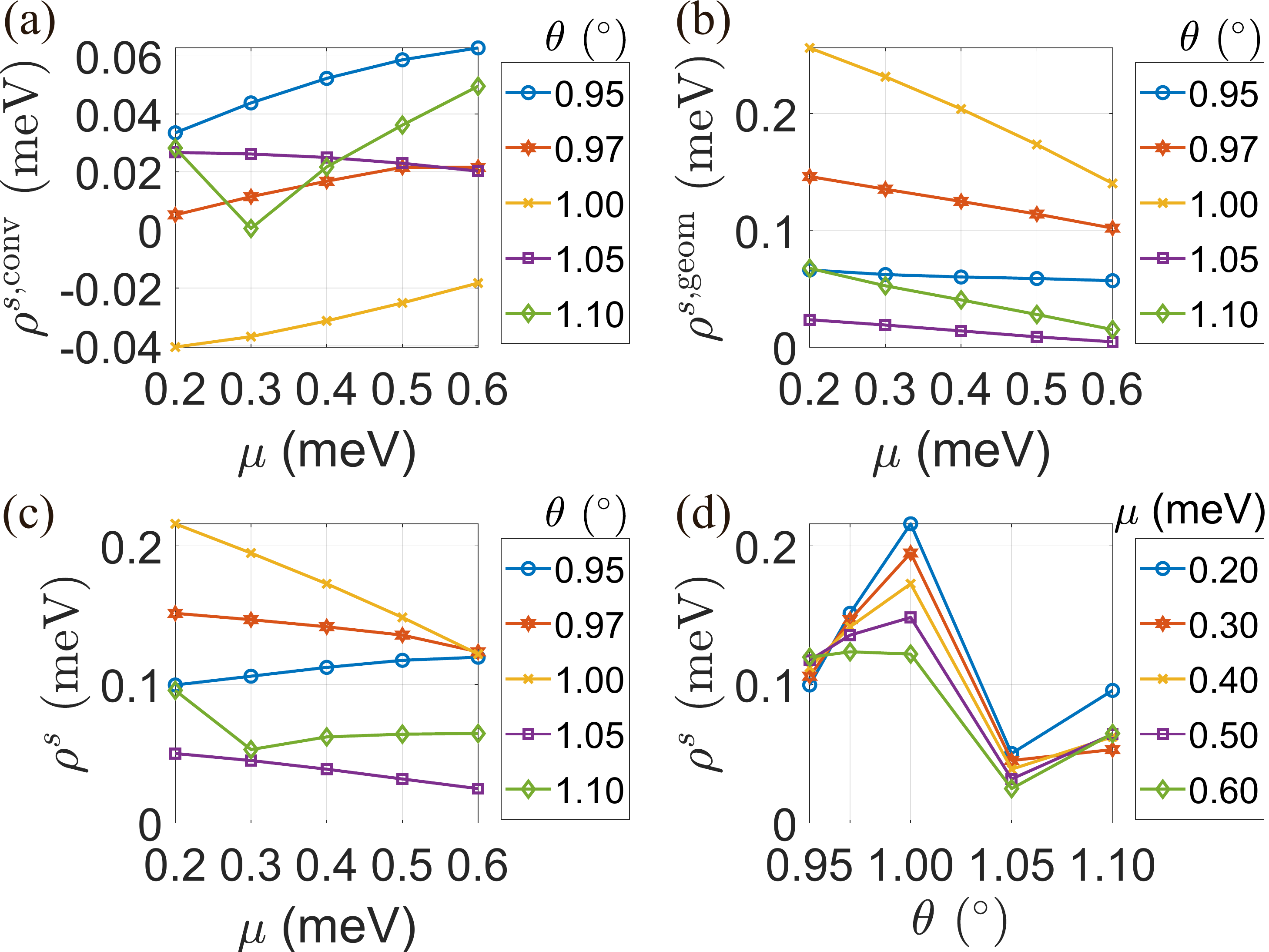}
  \caption{
(a) Conventional $\rho^{s,{\rm conv}}$, (b) geometric $\rho^{s,{\rm geom}}$, and (c) total stiffness $\rho^{s}$ as a function of $\mu$ for different values of $\theta$.
           (d)  $\rho^{s}$ vs. $\theta$ for different values of $\mu$.
         }
  \label{fig:rhos}
 \end{center}
\end{figure}

Fig~\ref{fig:rhos} shows how $\rho^{s,{\rm conv}}$, $\rho^{s,{\rm geom}}$
and $\rho^{s}$ depend on $\mu$
and $\theta$.
All the results were obtained for $\mathcal{T}=20$ mK $\ll T_c$.
We notice that  \rs does not
grow with $\mu$ contrary to the conventional
result $\rho_s\propto\mu$.
For $\theta=1.05\degree$, and $\theta=1.10\degree$,
$\rho^{s,{\rm conv}}$ and $\rho^{s,{\rm geom}}$ are comparable
and the relative weight changes with $\mu$.
For all the other twist
angles considered  $\rho^{s,{\rm geom}}$ is larger than $\rho^{s,{\rm conv}}$, regardless of $\mu$.

The results of Fig.~\ref{fig:Tc}~(a) show that the mean field critical temperature $T_c$ at  $\theta=1.00\degree$
is only slightly smaller than at $\theta=\theta_M$,
and therefore that, at the mean-field level,
double-layer TBLG with $\theta=1.00\degree$ is  a
very good candidate for the realization of an EC.
However, strikingly,  for $\theta=1.00\degree$
we find that $\rho^{s,{\rm conv}}$ for the EC is negative for all the values of $\mu$, see Fig.~\ref{fig:rhos}~(a) (this can happen because of the lack of particle-hole symmetry).
This result would lead us to conclude
that for $\theta=1.00\degree$
the EC is fragile against fluctuations and therefore not a stable ground state, despite the relatively large value of $T_c$.
This conclusion is reversed if one takes into account the geometric contribution
to \rhos, Fig.~\ref{fig:rhos}~(b): for $\theta=1.00\degree$
the $\rho^{s,{\rm geom}}$ is positive and much larger, in absolute value, than
$\rho^{s,{\rm conv}}$, guaranteeing the robust stability of the EC.
In fact, Figs.~\ref{fig:rhos}~(c),~(d) allow us to conclude that the EC is most stable for $\theta=1.00\degree$,
not for $\theta=\theta_M$ as one would infer from the mean-field results.

\begin{figure}[!!!t]
 \begin{center}
  \centering
  \includegraphics[width=\columnwidth]{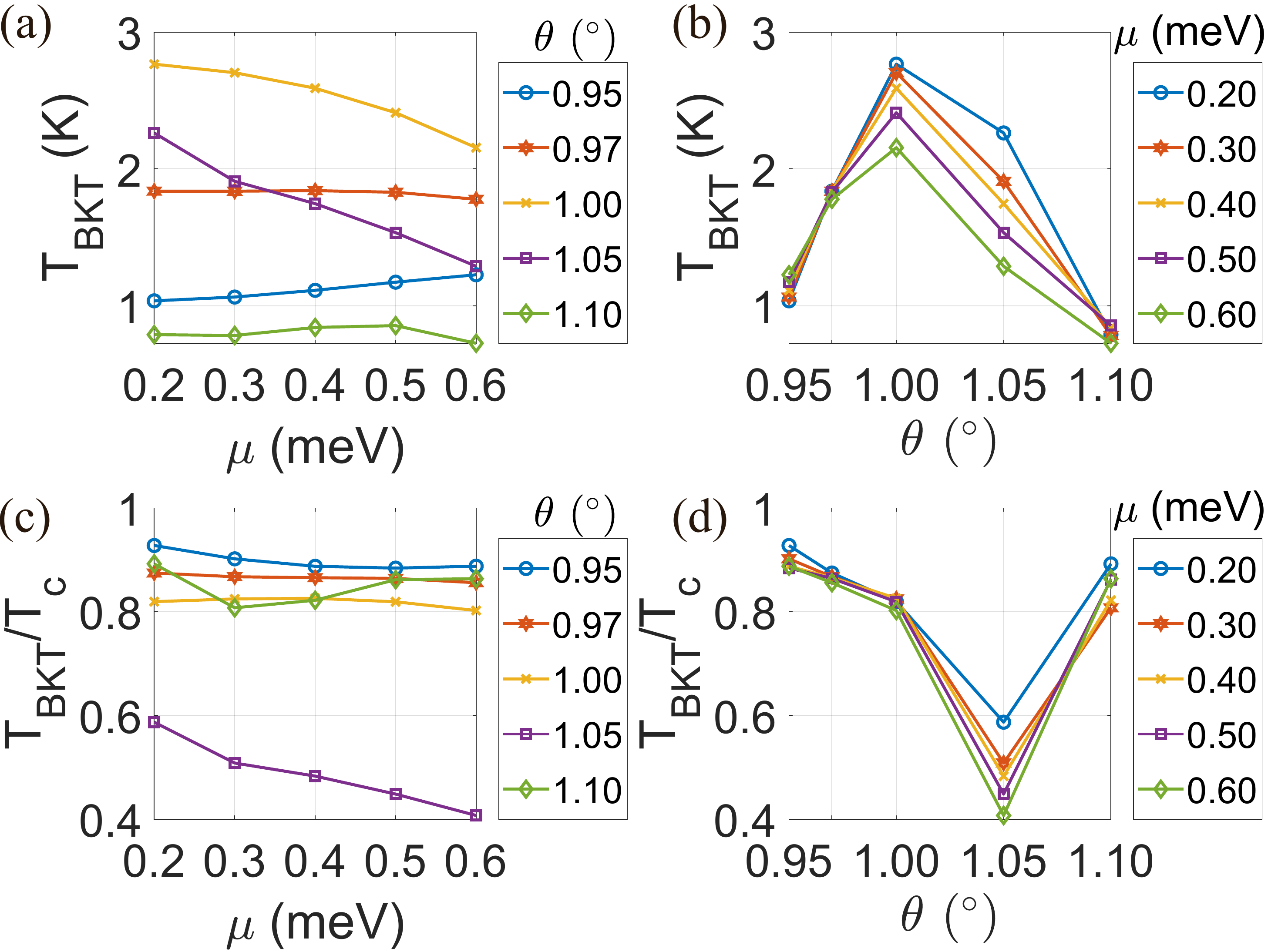}
  \caption{
           (a) $T_{\mathrm{BKT}}$ as a function of $\mu$ for different values of $\theta$.
           (b) $T_{\mathrm{BKT}}$ as a function of $\theta$ for different values of $\mu$.
           (c), (d) $T_{\mathrm{BKT}}/T_c$ as a function of $\mu$, $\theta$, respectively.
         }
  \label{fig:TKT}
 \end{center}
\end{figure}

The results of Fig.~\ref{fig:rhos}(c),(d) can be used to obtain $T_{\mathrm{BKT}}$,
via the equation $k_B T_{\mathrm{BKT}} = 2 \pi \rho^s[\Delta(T_{\mathrm{BKT}}),T_{\mathrm{BKT}}]$, where we have taken into account the valley and spin degeneracies.
For the dependence of $\Delta$ on $\mathcal{T}$
we can adopt the BCS scaling
$\Delta(\mathcal{T})=1.764 k_B T_c(1-\mathcal{T}/T_c)^{1/2}$, with $k_B$ the Boltzmann's constant.
The results for $T_{\mathrm{BKT}}$ are shown in Fig.~\ref{fig:TKT}.
From Fig.~\ref{fig:TKT}~(a),~(b) we see that, contrary to the mean-field results, the twist angle for which
the critical temperature $T_{\mathrm{BKT}}$ is largest is not $\theta_M$,
but $\theta=1.00\degree$, for all the values of $\mu$. Indeed
$T_{\mathrm{BKT}}$ at $\theta=1.00\degree$ is up to 50\% larger than at $\theta_M$.
This somewhat surprising result arises entirely from the
geometric contribution to \rs.
It is interesting to notice that, contrary to the conventional wisdom,
for some twist angles $T_{\mathrm{BKT}}$ decreases, rather than increasing, with $\mu$.
Such behavior is particularly marked for $\theta=1.00\degree$ and $\theta=\theta_M$, Fig.~\ref{fig:TKT}~(a),
due to the significant decrease of the geometric contribution to \rs, as seen in Fig.~\ref{fig:rhos}.
Figures~\ref{fig:TKT}~(c),~(d) show how the ratio $T_{\mathrm{BKT}}/T_c$ scales with $\mu$ and $\theta$, respectively.
It is particularly interesting to see that, for all values of $\mu$,
$T_{\mathrm{BKT}}/T_c$ is minimum at $\theta_M$.

In summary, we have studied the competition between OM, SC and EC phases as a function of the dopings of the layers  via  comprehensive mean-field calculations in double TBLG systems.
We have discussed the nature of the phase transitions,
and we have shown that for realistic interaction strengths the EC phase is favored when the TBLGs have sufficiently large and opposite dopings.
We  then studied the stiffness $\rho_s$ of the EC and demonstrated that the
quantum metric contribution to $\rho_s$ is essential to make \rs positive so that the EC is stable against fluctuations.
A ``conventional'' study of the EC's stability that does not include interbands terms
would lead to the conclusion that in flat-band double layers ECs can be unstable. However, we found that this conclusion is reversed if the interband terms responsible for the quantum metric of the flat bands are taken into account.
Finally, we
obtained $T_{\mathrm{BKT}}$ for the ECs and found that the largest $T_{\mathrm{BKT}}$ is realized not at the magic angle, $\theta=1.05\degree$, but at $\theta=1.00\degree$.
The results present a comprehensive and detailed
picture of the possible correlated states of double-twisted bilayer graphene,
and show the role played by the quantum metric on
the stability and $T_{\mathrm{BKT}}$  of the exciton condensate in double-twisted bilayer graphene
and so should constitute a useful guide to experimentalists studying the correlated phases of these novel systems.
In a more general context, our findings point to the importance of the quantum metric for the understanding of
the physics of ECs in flat band systems, including QH and moir\'e bilayers~\cite{Shi2021,Jang2021,Wang2021}.

\begin{acknowledgments}
X.H and E.R. acknowledge support from ARO (Grant No. W911NF-18-1-0290) and NSF (Grant No. DMR- 1455233). E.R. also thanks the Aspen Center for Physics, which is supported by NSF Grant No. PHY-1607611, and KITP, supported by  Grant No. NSF PHY1748958, where part of this work was performed. X.H acknowledge the hospitality of Hunan Normal University.
The authors acknowledge William \& Mary Research Computing for providing computational resources.
Part of the calculations were performed on
the Extreme Science and Engineering Discovery Environment (XSEDE)~\cite{xsede2014} Stampede2 at TACC through allocation TG-PHY210052.
The research was partially supported by the Foundation for Polish Science through the IRA Programme co-financed by EU within SG OP.
\end{acknowledgments}

%



\setcounter{equation}{0}
\setcounter{figure}{0}
\setcounter{table}{0}
\renewcommand{\theequation}{S\arabic{equation}}
\renewcommand{\thefigure}{S\arabic{figure}}
\renewcommand{\thetable}{S\arabic{table}}
\renewcommand{\bibnumfmt}[1]{[S#1]}
\renewcommand{\citenumfont}[1]{S#1}
\newcommand{\bk}{\boldsymbol\kappa}

\section{Supplemental material}

\section{I. Momentum space tight-binding model}

\begin{figure}[t]
\begin{center}
 \includegraphics[width=0.9\columnwidth]{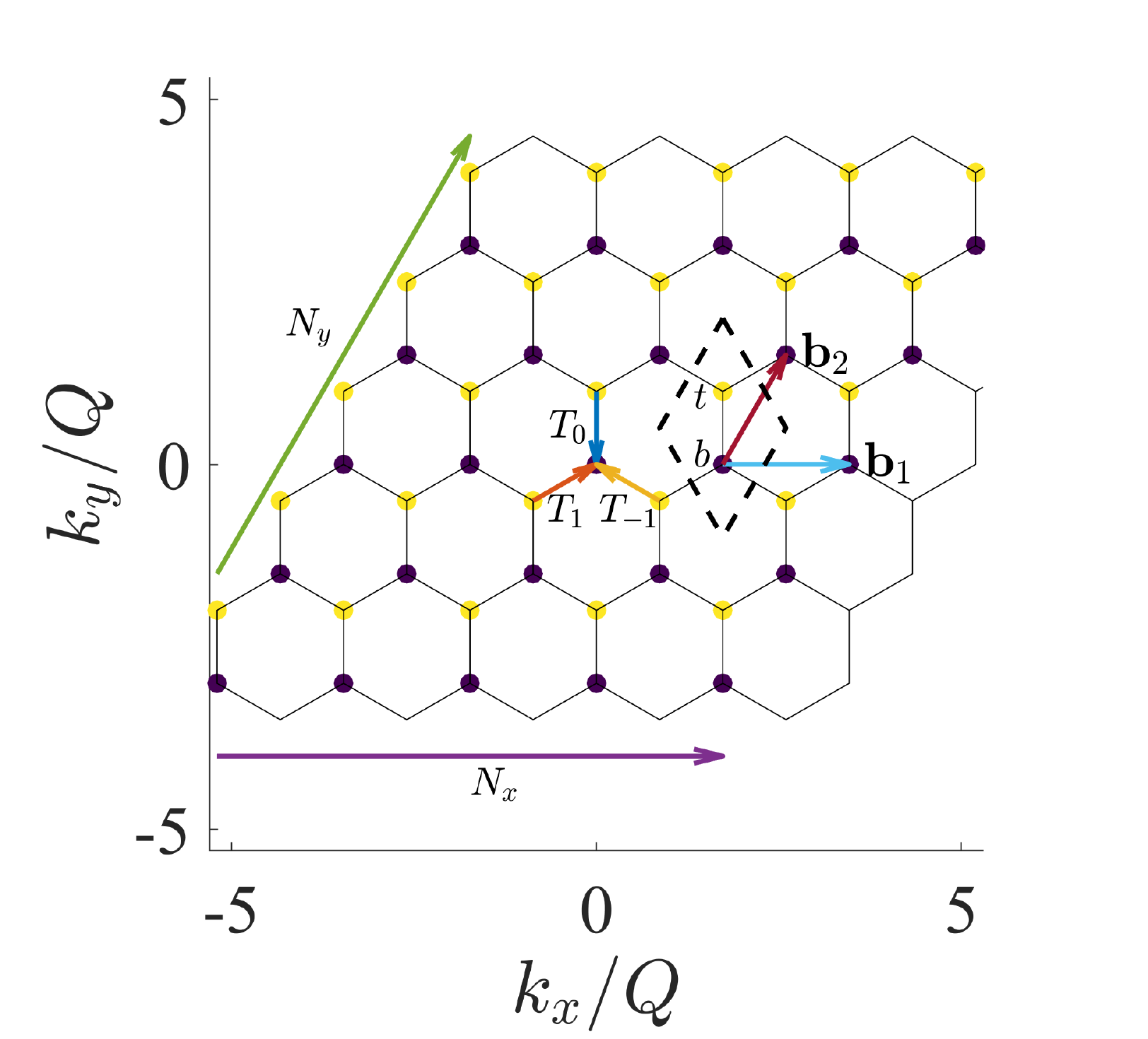}
 \caption{The Dirac points in each layer form a triangular lattice with
basis vectors $\bb_1$ and $\bb_2$. Because the momenta of the Dirac points in the two layers are shifted by ${\boldsymbol\beta}_+=(0,0)$ and ${\boldsymbol\beta}_-=(0,Q)$, we obtain a honeycomb lattice of Dirac points coupled by matrices  $T_j$  describing the tunneling between the layers. $N_x$ and $N_y$
are the number of primitive cells, along the ${\bf b}_1$, ${\bf b}_2$ direction, respectively, forming the lattice.
}
\label{FigS:BZ}
\end{center}
\end{figure}

For all the calculations of TBLG in this paper, we adopt the model introduced by Bistrizer-MacDonald~\cite{Bistritzer2011aS,Wu2018S}, which is an effective tight-binding model in momentum-space (see Fig.~\ref{FigS:BZ}). The Dirac points in each layer form a triangular lattice with the reciprocal basis vectors
\begin{equation}
\bb_1=(\sqrt{3}Q, 0),
\end{equation}
and
\begin{equation}
\bb_2=(\sqrt{3}Q/2, 3Q/2).
\end{equation}
where $Q=8\pi\sin(\theta/2)/3a_0$, $a_0$ is the lattice constant for graphene and $\theta$ is the twist angle. The Dirac points in the two layers are shifted relative to each other by ${\boldsymbol\beta}_+=(0,0)$ and ${\boldsymbol\beta}_-=(0,Q)$, so that we end up with an effective honeycomb lattice. The onsite block at each lattice site $\bb$, corresponding to the $\KK$ valley of the graphene sheets,  is described by the
Hamiltonian
\begin{equation}
 H^{(t/b)}_{\KK \bb}(\kk)= e^{\mp i\frac{\theta}{4}\sigma_z}[\hbar v_F(\kk-{\boldsymbol\kappa}_{\bb \pm})\cdot\ssigma -\mu\sigma_0]e^{\pm i\frac{\theta}{4}\sigma_z},
 \label{eq:H_slg}
\end{equation}
where the first (second)
sign  is associated to the bottom (top), i.e. $b$, ($t$), graphene sheet,
$\sigma_i$ are Pauli matrices in sublattice $A/B$ space, $v_F=10^6$~m/s is the Fermi velocity in graphene,
$\kk$ is the electron wave vector (measured from $\KK$), ${\boldsymbol\kappa_{\bb\pm}}=\bb+{\boldsymbol\beta_{\pm}}$,
and $\mu$ is the chemical potential.
The matrices
$T_j=w[\tau_0+\cos(2\pi j/3)\tau_x+\sin(2\pi j/3)\tau_y]$~\cite{Bistritzer2011aS, Wu2018S},
with $j=-1,0,1$, describe tunneling processes between
nearest neighbors on the lattice sites built in momentum space,
see Fig.~\ref{FigS:BZ}.
$H_{\KK'}$ is obtained from $H_{\KK}$ via time reversal.
We assume $w=118$~meV~\cite{Jung2014S,Carr2018S}.

For the determination of the phase diagram and the stiffness calculations, we use lattices with $N_x=11$ and $N_y=11$ primitive cells
along the  $\bb_1$ and $\bb_2$ directions, respectively,
resulting in $2N_xN_y=242$ momentum-space lattice points.
To solve the non-linear gap equations we consider the momentum-space unit-cells with $|\bb|\leq 5|\bb_1|$, which corresponds to $2\times91=182$ momentum-space lattice points.
We have verified that both sets of  momentum-space lattice points give the same results and that are sufficiently large that the results do not change if the number of momentum-space lattice points is further increased.

\section{II. Justification of local-interaction approximation}
In this section we justify the local-interaction approximation that we have used to obtain the mean-field equations.
Let's consider the exciton order parameter $\Delta(\mathbf{k})$ assuming that the spin- and valley-rotation symmetries are satisfied, so that we can neglect these degrees of freedom in our analysis. Then the mean field Hamiltonian projected to the lowest energy bands of the two TBLGs (denoted $U$ and $L$) is given by
\begin{equation}
H_{\rm MF}=\sum_{k} c^\dag_\mathbf{k} \begin{pmatrix} \xi_0(\mathbf{k})+\xi_z(\mathbf{k}) & \Delta(\mathbf{k}) \\ \Delta(\mathbf{k}) & \xi_0(\mathbf{k})-\xi_z(\mathbf{k}) \end{pmatrix} c_\mathbf{k},
\end{equation}
where $c^\dag_\mathbf{k}=(c^\dag_{U, \mathbf{k}} \ c^\dag_{L, \mathbf{k}})$, $\xi_0(\mathbf{k}) \pm \xi_z(\mathbf{k})$ are the normal state dispersions in the U/L layers,
\begin{eqnarray}
\Delta(\mathbf{k})&=&\sum_{\mathbf{k}'} V(\mathbf{k}, \mathbf{k}') \frac{\Delta(\mathbf{k}')}{2 \sqrt{\xi^2_z(\mathbf{k}')+\Delta^2(\mathbf{k}')}}\nonumber\\
&&\bigg[ n_F\big(-\sqrt{\xi^2_z(\mathbf{k}')+\Delta^2(\mathbf{k}')}+\xi_0(\mathbf{k})\big)\nonumber\\
&&-n_F\big(\sqrt{\xi^2_z(\mathbf{k}')+\Delta^2(\mathbf{k}')}+\xi_0(\mathbf{k})\big) \bigg],
\end{eqnarray}
and $V(\mathbf{k}, \mathbf{k}')$ describes the electron-electron interactions projected to the lowest bands.  Close to the critical temperature this gap equation can be linearized giving
\begin{eqnarray}
\Delta(\mathbf{k})&=&\sum_{\mathbf{k}'} V(\mathbf{k}, \mathbf{k}') \frac{\Delta(\mathbf{k}')}{2 |\xi_z(\mathbf{k}')|}\bigg[ n_F\big(-|\xi_z(\mathbf{k}')|+\xi_0(\mathbf{k})\big)\nonumber\\
&&-n_F\big(|\xi_z(\mathbf{k}')|+\xi_0(\mathbf{k})\big) \bigg].
\end{eqnarray}

In general, the exciton order parameter is momentum-dependent due to the dispersion of the bands. However, if we assume that the dispersions of the bands are approximately flat we can neglect the momentum-dependence of $\Delta(\mathbf{k})$. Then the non-linear and linear gap equations take the forms
\begin{eqnarray}
\Delta&=&V_{\rm eff}\frac{\Delta}{2 \sqrt{\xi^2_z+\Delta^2}}\bigg[ n_F\big(-\sqrt{\xi^2_z+\Delta^2}+\xi_0\big)\nonumber\\
&&-n_F\big(\sqrt{\xi^2_z+\Delta^2}+\xi_0\big) \bigg],
\end{eqnarray}
and
\begin{equation}
\Delta= V_{\rm eff} \frac{\Delta}{2 |\xi_z|}\bigg[ n_F\big(-|\xi_z|+\xi_0\big)-n_F\big(|\xi_z|+\xi_0\big) \bigg],
\end{equation}
respectively, where
\begin{equation}
V_{\rm eff} = \sum_{\mathbf{k}'} V(\mathbf{k}, \mathbf{k}').
\end{equation}
This means that in both cases we can replace the interactions with a momentum-independent effective interaction strength $V_{\rm eff}$ so that  neither the critical temperature obtained from the linearized gap equations, nor the order parameter obtained from the full non-linear gap equations is affected by the detailed momentum dependence of the interactions.

Following the analysis above, it can be shown that also to calculate the
superconducting and magnetic order parameters the interaction is well approximated
by an effective local interaction.
In general, in systems with sufficiently flat bands,
local and long-range interactions will yield similar results. This has been utilized in numerous works related to symmetry-broken states in quantum Hall bilayers, see e.g.~Ref.~\cite{Moon1995S,Pikulin16S},
where it was shown that the long-range Coulomb interactions can be replaced by effective local interactions with interaction strength $V_c \sim e^2/(4 \pi \epsilon \epsilon_0 l_B)$, where $l_B$ is the magnetic length describing the distance between the particles. Moreover, a similar projection of the interactions to the flat bands has been used to replace the momentum-dependent Coulomb repulsion, and the phonon-mediated attraction, with momentum-independent effective interaction strengths in other flat-band systems~\cite{Ojajarvi2018S,Pikulin2021S}.

Given that in TBLG the low energy bands are very flat the local-interactions approximation is quite accurate.
It might be worth pointing out that our results, obtained using the  local-interactions approximation,
are in agreement with the mean-field results obtained in Ref.~\cite{Xie2020S},
where the long-range Coulomb interactions were used.
Our results are also qualitatively consistent with the experimentally observed doping dependence of the competition of the superconducting and correlated insulating phases. One situation when, for TBLG, the local-interaction approximation would not be valid is the case when the translational symmetry of the moir\'e superlattice is spontaneously broken due to a finite momentum pairing, or charge density wave, or spin density wave order. In these cases, the comparison of the energies of the candidate ground states would require us to take into account the momentum-dependence of the interactions. However, to our knowledge, there is no evidence of spontaneously broken translational symmetry of the moir\'e superlattice in TBLGs so far, and that is not a situation that we consider in the current work.
In quantum Hall exciton condensates, it is known that the order parameters can spontaneously break the translational symmetry if the distance between the layers is larger than the distance between the particles within the layer~\cite{Cote1992S}.
In our case, the distance between the layers can be just few atomic constants because the insulating layer can be made out of h-BN or WSe$_2$ having a very large insulating gap. Therefore, we do not expect that phases with spontaneously broken translational symmetry will appear in our system.


\section{III. Phase diagram for an ideal flat-band model}

To get insights into the form of the phase
diagram we can assume the bands to be completely flat.
This is a rough approximation, but it is sufficient to
identify the main qualitative features of the phase diagram.
Almost flat bands can not only be obtained in systems formed by two two-dimensional crystal stacked with a relative twist,
but can also be realized in optical lattices~\cite{Luo2020S}.
In the limit in which the bands are completely flat
the non-interacting Hamiltonian can be written as
\begin{eqnarray}
H_{\rm flat}&=&-\mu_U\sum_{ks} c^\dagger_{\kk s U}c_{\kk s U}-\mu_L\sum_{k s} c^\dagger_{\kk s L}c_{\kk s L},
\end{eqnarray}
where $s$ represents a spin or orbital degree of freedom.

As discussed in the main text: (i) close to half-filling $\mu_U=\mu_L=0$ we expect an intralayer correlated state, (ii) for  $\mu_U=\mu_L$ and $|\mu_U|$ sufficiently large we expect a superconducting state, (iii) for $\mu_U=-\mu_L$, and $|\mu_U|$ sufficiently large, we expect an exciton condensate state [Fig.~1(b) in the main text].
In this section we show that this expectation can be confirmed with a simple analysis based on the structure of the gap equations resembling the approach used in Ref.~\cite{Black17S}.
We describe the gap equations for each type of symmetry-broken state separately, and then use them to study the competition between the phases.

\noindent
\begin{figure}[!!!t]
 \begin{center}
  \centering
  \includegraphics[width=0.6\columnwidth]{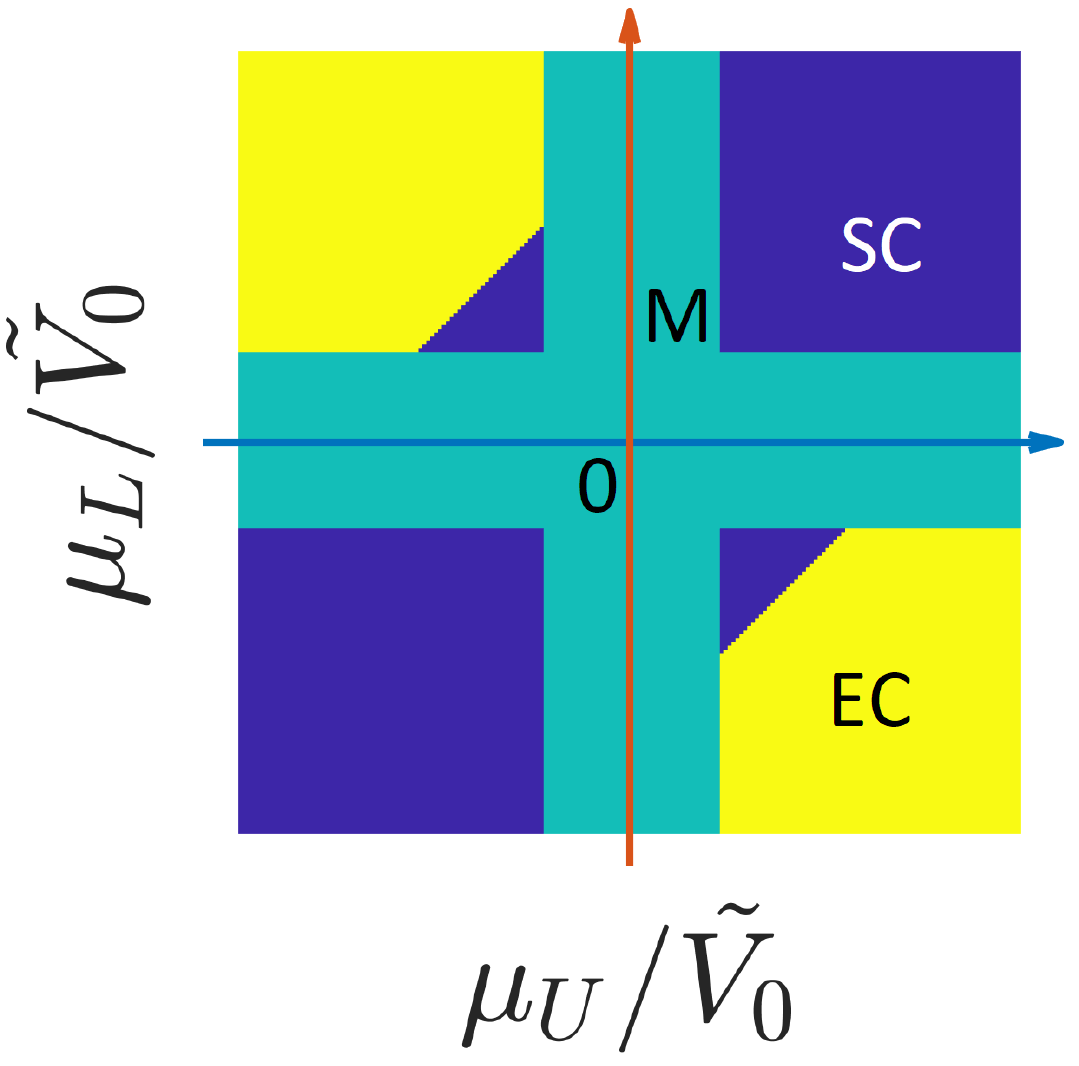}
  \caption{Phase diagram of a double layer system with ideal flat bands, here $\tilde{V}_0$: effective interaction strength for the correlated insulator states.
         }
  \label{fig:flatpd}
 \end{center}
\end{figure}

The gap equation for $s-$wave
superconductivity in each layer is (we can assume that superconducting pairing in each layer occurs independently)
\begin{eqnarray}
\Delta_{{\rm SC}, T}&=&\frac{V_{\rm SC}}{2} \frac{\Delta_{{\rm SC}, T}}{\sqrt{\mu_T^2+\Delta^2_{{\rm SC}, T}}} \bigg[ n_F\big(- \sqrt{\Delta_{{\rm SC}, T}^2+\mu_T^2} \big)\bigg.\nonumber\\&&\bigg.- n_F\big(\sqrt{\Delta_{{\rm SC}, T}^2+\mu_T^2}\big)\bigg]\nonumber\\
&=&\frac{V_{\rm SC}}{2} \frac{\Delta_{{\rm SC}, T}}{\sqrt{\mu_T^2+\Delta^2_{{\rm SC}, T}}} \tanh(\beta \sqrt{\Delta_{{\rm SC}, T}^2+\mu_T^2}/2),\nonumber\\
\end{eqnarray}
where $n_F(E)$ is the Fermi function, $V_{\rm SC}$ is the effective interaction strength for superconductivity obtained by projecting the true interactions into the eigenstates of the flat-band system, $\mu_T$ ($T=U,L$) is the chemical potential in each layer, $\Delta_{{\rm SC}, T}$ is the superconducting order parameter in each layer, $\beta=1/(k_BT_0)$ and $T_0$ is the temperature.
The critical temperature $T_{{\rm SC}, T, c}$ for the superconducting state is given by
\begin{equation}
k_B T_{{\rm SC}, T, c} = \frac{\mu_T}{2 \ {\rm arctanh} \big(2 \mu_{T}/V_{\rm SC} \big)}. \label{Tc-SC}
\end{equation}
The largest critical temperature $k_B T_{{\rm SC}, T, c}=V_{\rm SC}/4$ is obtained for $\mu_T=0$.

For an intralayer symmetry-broken state (say for magnetism) the gap equation is
\begin{eqnarray}
m_{T}&=&\frac{\tilde{V}_0}{2} \bigg[n_F(-m_T-\mu_T) - n_F(m_T-\mu_T) \bigg]\nonumber\\
&=&\frac{\tilde{V}_0}{2}\, \frac{\sinh\beta m_{T}}{\cosh \beta\mu_T + \cosh \beta m_{T}},
\end{eqnarray}
where $\tilde{V}_0$ is the effective interaction strength for intralayer interactions and $m_{T}$ is the order parameter of the intralayer symmetry-broken state.   The critical temperature of the  intralayer correlated state is determined by the equation
\begin{equation}
k_B T_{0,T, c} \bigg[\cosh(\mu_T/k_B T_{0, T, c} ) + 1 \bigg]  =  \frac{\tilde{V}_0}{2}. \label{Tc-0}
\end{equation}
The largest critical temperature $k_B T_{0, T, c}=\tilde{V}_{0}/4$ is obtained for $\mu_T=0$.

For an exciton condensate state the gap equation is
\begin{eqnarray}
\Delta_{\rm EC}&=&\frac{V_{\rm EC}}{2} \frac{\Delta_{\rm EC}}{\sqrt{\mu_{as}^2+\Delta^2_{\rm EC}}} \bigg[ n_F\bigg(- \sqrt{\Delta_{\rm EC}^2+\mu_{as}^2} - \mu_s \bigg)\bigg.\nonumber\\
&&\bigg.- n_F\bigg(\sqrt{\Delta_{\rm EC}^2+\mu_{as}^2}-\mu_s \bigg)\bigg] \nonumber\\
&=& \frac{V_{\rm EC}}{2} \frac{\Delta_{\rm EC}}{\sqrt{\mu_{as}^2+\Delta^2_{\rm EC}}}\nonumber\\
&&\times \frac{\sinh \big(\beta \sqrt{\mu_{as}^2+\Delta^2_{\rm EC}}\big)}{\cosh \beta\mu_s + \cosh \big(\beta \sqrt{\mu_{as}^2+\Delta^2_{\rm EC}}\big)},
\end{eqnarray}
where $V_{\rm EC}$ is the effective interaction strength for exciton condensation, $\mu_s=(\mu_U+\mu_L)/2$ and $\mu_{as}=(\mu_U-\mu_L)/2$. If we have balanced electron and hole densities $\mu_U=-\mu_L$ the gap equation for the exciton condensation has the same form as in the case of superconductivity. If the layers have equal densities of similar type of carriers $\mu_U=\mu_L$ the gap equation for the exciton condensation takes the same form as in the case of intralayer correlated state. The critical temperature for the exciton condensation is determined by the equation
\begin{equation}
\frac{\sinh \big(\mu_{as}/k_B T_{{\rm EC},c}\big)}{\cosh(\mu_s/k_B T_{{\rm EC},c}) + \cosh \big(\mu_{as}/k_BT_{{\rm EC}, c} \big)} = \frac{2 \mu_{as}}{V_{\rm EC}}. \label{Tc-EC}
\end{equation}
If $\mu_s=0$ we obtain
\begin{equation}
k_B T_{{\rm EC},c} =  \frac{\mu_{as}}{2 \ {\rm arctanh}(2 \mu_{as}/V_{\rm EC})}
\end{equation}
resembling the equation for critical temperature of superconductivity but with chemical potential replaced by the layer asymmetric chemical potential $\mu_{as}=(\mu_U-\mu_L)/2$.
If $\mu_{as}=0$ we obtain
\begin{equation}
 k_B T_{{\rm EC},c} \bigg[\cosh(\mu_s/k_B T_{{\rm EC},c}) + 1 \bigg] =\frac{V_{\rm EC}}{2}
\end{equation}
resembling the equation for critical temperature of intralayer correlated state but with chemical potential replaced by the layer symmetric chemical potential $\mu_{s}=(\mu_U+\mu_L)/2$.
The largest critical temperature $k_B T_{{\rm EC}, c}=V_{\rm EC}/4$ is obtained for $\mu_U=\mu_L=0$.

To describe the competition between different phases as a function of $\mu_U$ and $\mu_L$ we use the following procedure: For each $\mu_U$ and $\mu_L$ we solve the critical temperatures for the superconductivity from Eq.~(\ref{Tc-SC}) (taking the larger value from $T_{{\rm SC}, U, c}$ and $T_{{\rm SC}, L, c}$), intralayer correlated states from Eq.~(\ref{Tc-0}) (taking the larger value from $T_{0,U, c}$ and $T_{0,L, c}$) and exciton condensation from Eq.~(\ref{Tc-EC}). We assume that the phase which is realized at each $\mu_U$ and $\mu_L$ is the the one with the largest critical temperature.

We assume that the interaction constants satisfy the realistic hierarchy $\tilde{V}_0>V_{\rm EC}\sim V_{\rm SC}$. Knowing that superconductivity can be easily realized in twisted bilayer graphene we can expect $V_{\rm SC}$ to be not much smaller than $\tilde{V}_0$.
As an example, Fig.~\ref{fig:flatpd} shows the phase diagram obtained assuming $\tilde{V}_0=162$meV,
$V_{\rm EC}=0.8\tilde{V}_0$ and $V_{\rm SC}=0.6\tilde{V}_0$.
We see that in this case, as expected, we obtain the phase diagram discussed at the beginning of this section. We emphasize that the overall structure of this phase diagram is a universal consequence of the structure of the gap equations, and arises independently of the details of the microscopic model.

\section{IV. The determination of the phase diagram from the linearized gap equations}

In this section, we describe the approach used to obtain the phase diagram using the linearized gap equations obtained using the full TBLG bands, and provide the full expression of such equations.
As in the main text, we consider three different phases: exciton condensate (EC), superconductivity (SC), and orbital magnetism (OM).

To obtain the phase diagram, for each pair of
values in the $(\mu_U, \mu_L)$ plane we calculate the transition temperature ($T_c$) for each phase, and then identify the phase of the ground state as the one with the largest $T_c$.
As we discussed in the next section, we have verified
for several sample points in the $(\mu_U, \mu_L)$ plane
that the phase with the largest $T_c$ is the ground state phase (i.e. the phase at zero temperature) by solving the full non-linear gap equation, allowing also for the possibility of coexistence of different phases.

For the EC phase the linearized gap equation takes the form:
\begin{equation}
\Delta^{\rm EC}_{\bb_1 l_1\sigma_1 l'_1\sigma'_1}=\sum_{b_2 l_2\sigma_2 l'_2 \sigma'_2}\chi_{{\bf b}_1{\bf b}_2}^{l_1\sigma_1 l'_1\sigma'_1;l_2\sigma_2 l'_2\sigma'_2} \Delta^{\rm EC}_{\bb_2 l_2\sigma_2 l'_2\sigma'_2},
\label{eq:LOPec}
\end{equation}
where
\begin{eqnarray}
\chi_{{\bf b}_1{\bf b}_2}^{l_1\sigma_1 l'_1\sigma'_1;l_2\sigma_2 l'_2\sigma'_2}&=&\frac{V_{\rm EC}}{\mathcal{A}}\sum_{\substack{n_1 n_2\\b b' q}}\frac{n_F[\xi_{n_{2L}}({\bf q})]-n_F[\xi_{n_{1 U}}({\bf q})]}{\xi_{n_{1U}}({\bf q})-\xi_{n_{2L}}({\bf q})}\nonumber\\
&&u_{n_1 \bb_1+\bb l_1\sigma_1 U}(\qq)u_{n_2 \bb l'_1\sigma'_1 L}^*(\qq)\nonumber\\
&&u^*_{n_1 \bb'+\bb_2 l_2\sigma_2 U}(\qq)
u_{n_2 \bb' l'_2\sigma'_2 L}(\qq),
\end{eqnarray}
is the pairing susceptibility,
$V_{\rm EC}$ is the interaction strength for the EC phase, $\mathcal{A}$ is the area of the sample, $u_{n\bb l\sigma T}(\qq)$ is the component of the non-interacting wave function for a single upper ($T=U$) or lower ($T=L$) TBLG with reciprocal basis vector $\bb$, layer $l$, sublattice $\sigma$, and band index $n$. Here $\xi_{U,L}({\bf q})=\epsilon_\qq-\mu_{U,L}$, with $\epsilon_\qq$ the non-interacting energy eigenvalue.

For the superconducting case, we assume the pairing to be $s-$wave, and sublattice independent so that for each TBLG we have the linearized gap equation~\cite{Wu2018S}:
\begin{equation}
\Delta_{\bb l}^{\rm SC}=\sum_{\bb' l'}\chi_{\bb l,\bb' l'}^{\rm SC}\Delta_{\bb' l'}^{\rm SC},
\end{equation}
with
\begin{eqnarray}
\chi_{\bb l,\bb' l}^{\rm SC}&=&\frac{V_{\rm SC}}{2\mathcal{A}}\sum_{nmq}\frac{1-n_F(\xi_{n\qq})-n_F(\xi_{m\qq})}{\xi_{n\qq}+\xi_{m\qq}}\nonumber\\&&[\langle u_{n\qq}|u_{m\qq}\rangle_{\bb l}]^*
\langle u_{n\qq}|u_{m\qq}\rangle_{\bb' l'},
\end{eqnarray}
where the 2 in front of $\mathcal{A}$ is due to the summation over sublattices in the $\langle \rangle$ expression, and $n,m$ are band indexes~\cite{Wu2018S}.

Previous work \cite{Xie2020S}
strongly suggests that the orbital magnetic phase energetically most favored has order parameter
$\Delta_{OM}=\langle\Psi_{GS}|\hat{\sigma}_z|\Psi_{GS}\rangle$
where $|\Psi_{GS}\rangle$ is the ground-state and $\hat{\sigma}_z$ is the $z$ Pauli matrix in sublattice space.
For such order parameter we obtain the following linearized-gap equation
\begin{equation}
\Delta_{\bb l}^{\rm OM}=\sum_{b'l'}\chi^{\rm OM}_{{\bb l},{\bb' l'}}\Delta_{\bb' l'}^{\rm OM},
\end{equation}
with
\begin{eqnarray}
 &&\chi^{\rm OM}_{\bb l,\bb' l}=\frac{V_{\rm OM}}{\mathcal{A}}\sum_{nmq}\frac{n_F(\xi_{n\qq})-n_F(\xi_{m\qq})}{\xi_{m\qq}-\xi_{n\qq}}\nonumber\\
&&[\langle u_{n\qq A}|\hat{\sigma}_z|u_{m\qq A}\rangle_{\bb l}]^*
\langle u_{n\qq A}|\hat{\sigma}_z|u_{m\qq A}\rangle_{\bb' l'}.
\label{eq.chi_om}
\end{eqnarray}
The factor 2 in front of $\mathcal{A}$ on the r.h.s. of Eq.~\ceq{eq.chi_om} is cancelled by the factor 2 in front of $V_{\rm OM}$ due to the spin degeneracy.

To obtain the linear response functions we use
$60\times60$ k-points to calculate the sum in momentum space. We have verified that further increasing the number of k-ponts does not affect the phase diagram.

\begin{figure}[htbp]
\begin{center}
\includegraphics[width=0.65\columnwidth]{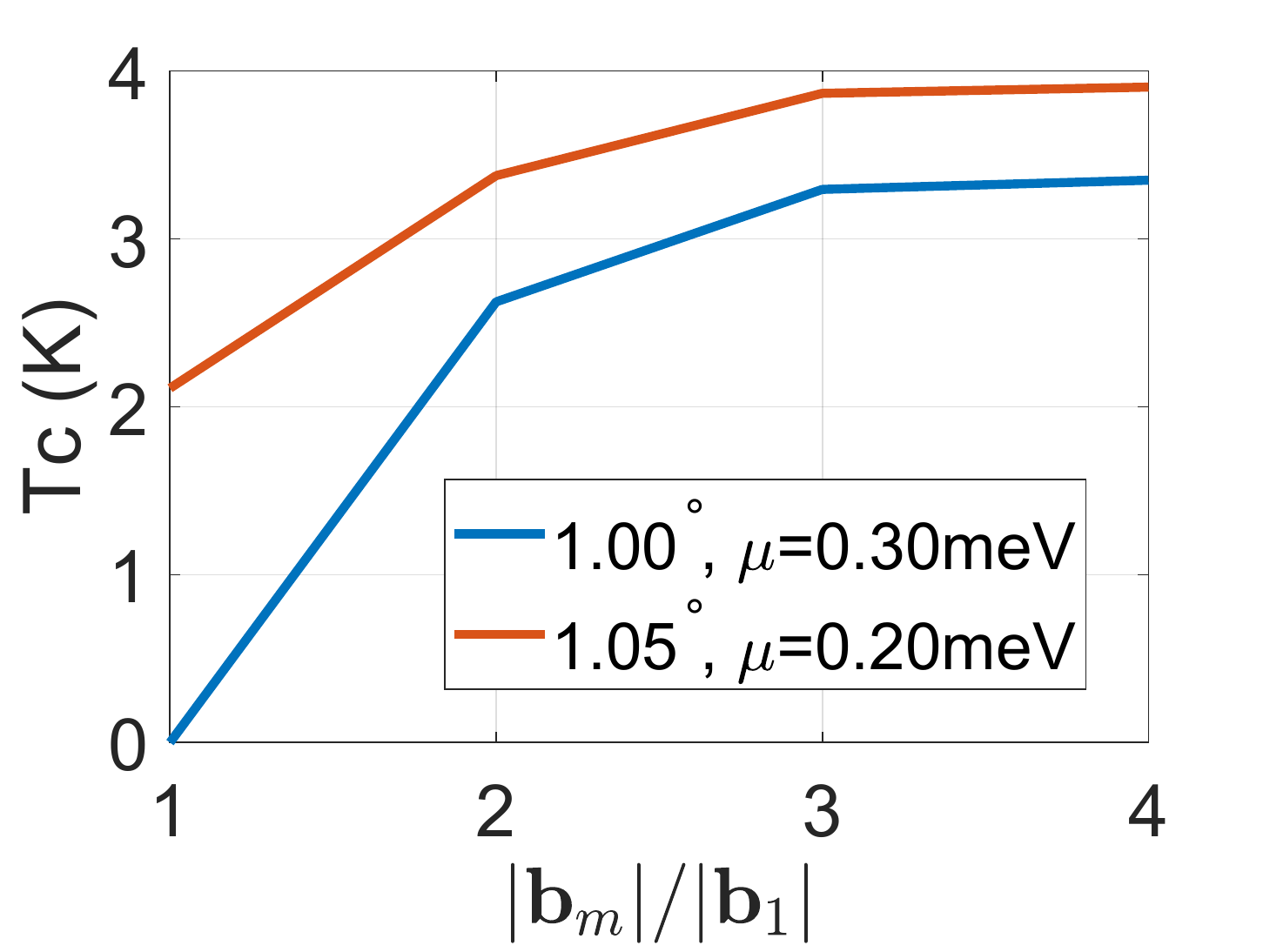}
\caption
{$T_c$ for the EC phase as a function of the magnitude of the largest $\bb$ vectors kept in solving the linearized gap equation.
The numbers of $\bb$ vectors are respectively 7, 19, 37, 61 with $|\bb_m/\bb_1|$ increased from 1 to 4.
}
\label{FigS:Tcb}
\end{center}
\end{figure}

The number of $\bb$ vectors necessary to obtain accurate estimates of $T_c$ depends on the order parameter considered.
For superconductivity it is sufficient to keep all the $\bb$ vectors with magnitude no larger than $2|\bb_1|$, which corresponds to 19 $\bb$ vectors~\cite{Wu2018S}.

For the EC order parameters we find that one needs to keep all the $\bb$ vectors with $|\bb|\leq 3 |\bb_1|$, i.e. a total of 37 $\bb$ vectors.
This is shown in Fig.~\ref{FigS:Tcb},
and is consistent with the fact
that the EC phase is characterized by a multi-component order parameter
requiring more momentum states to accurately describe it. Similar arguments and results apply to the OM phase. Consequently, all the results presented in the main text, and in the remainder, were obtained by keeping all the $\bb$ with $|\bb|\leq 4|\bb_1|$ resulting in a total of 61 $\bb$ vectors.

\noindent
\begin{figure}[!hbtp]
 \begin{center}
  \centering
  \includegraphics[width=\columnwidth]{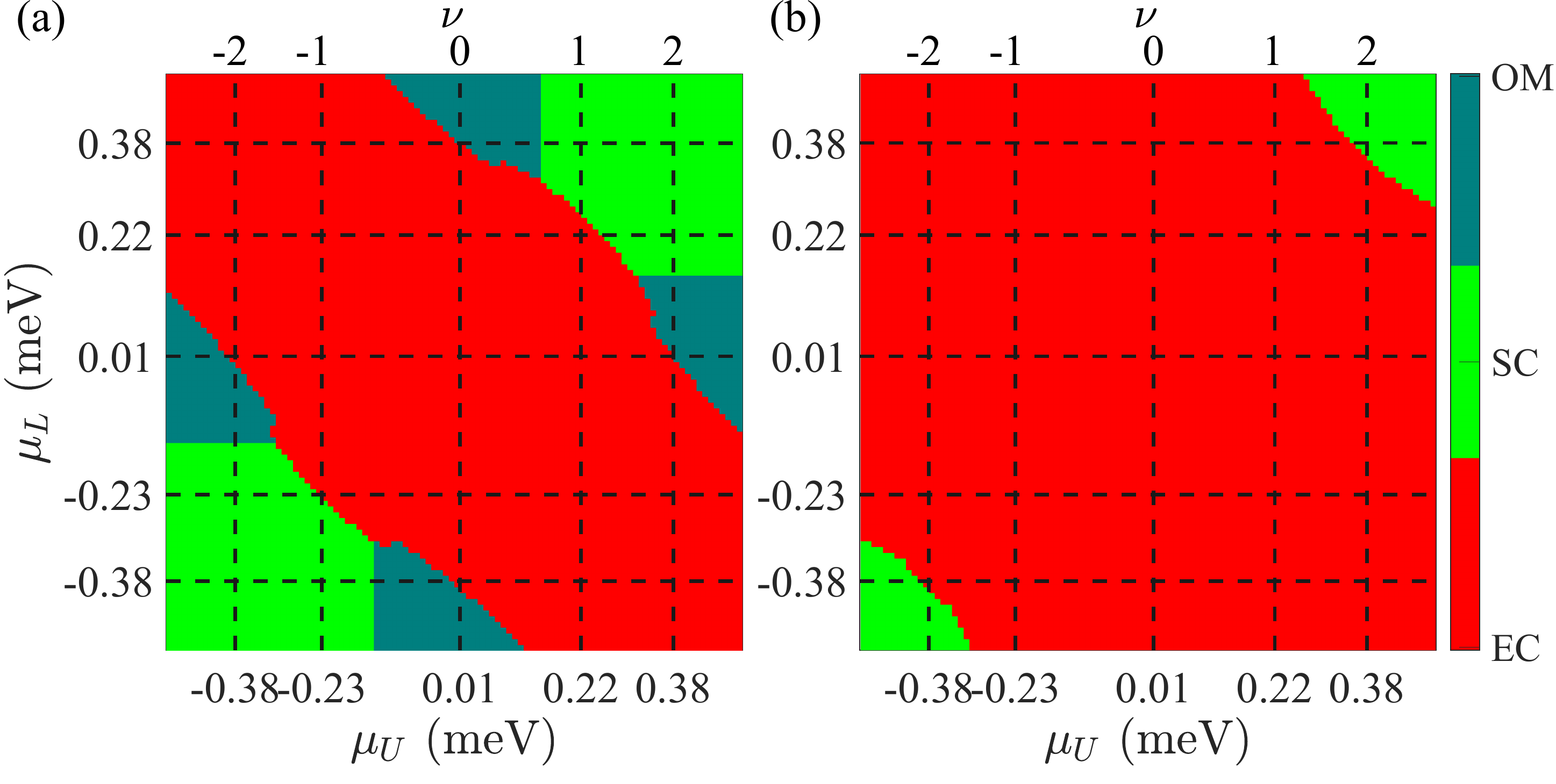}
  \caption{
  Effect of $V_{EC}$ on the phase diagram.
  For both panels:
  $\theta=1.00^\circ$, $V_{\rm SC}=75$~meV$\cdot$nm$^2$, and $V_{\rm OM}=130$~meV$\cdot$nm$^2$, the same values used for Fig.~1~(b) in the main text. However for (a) $V_{\rm EC}=80~$meV$\cdot$nm$^2$, and for (b) $V_{\rm EC}=100~$meV$\cdot$nm$^2$, whereas in Fig.~1~(b) in the main text $V_{\rm EC}=60~$meV$\cdot$nm$^2$.
         }
  \label{fig:pdV80V100}
 \end{center}
\end{figure}

\noindent
\begin{figure}[hbtp]
 \begin{center}
  \centering
  \includegraphics[width=\columnwidth]{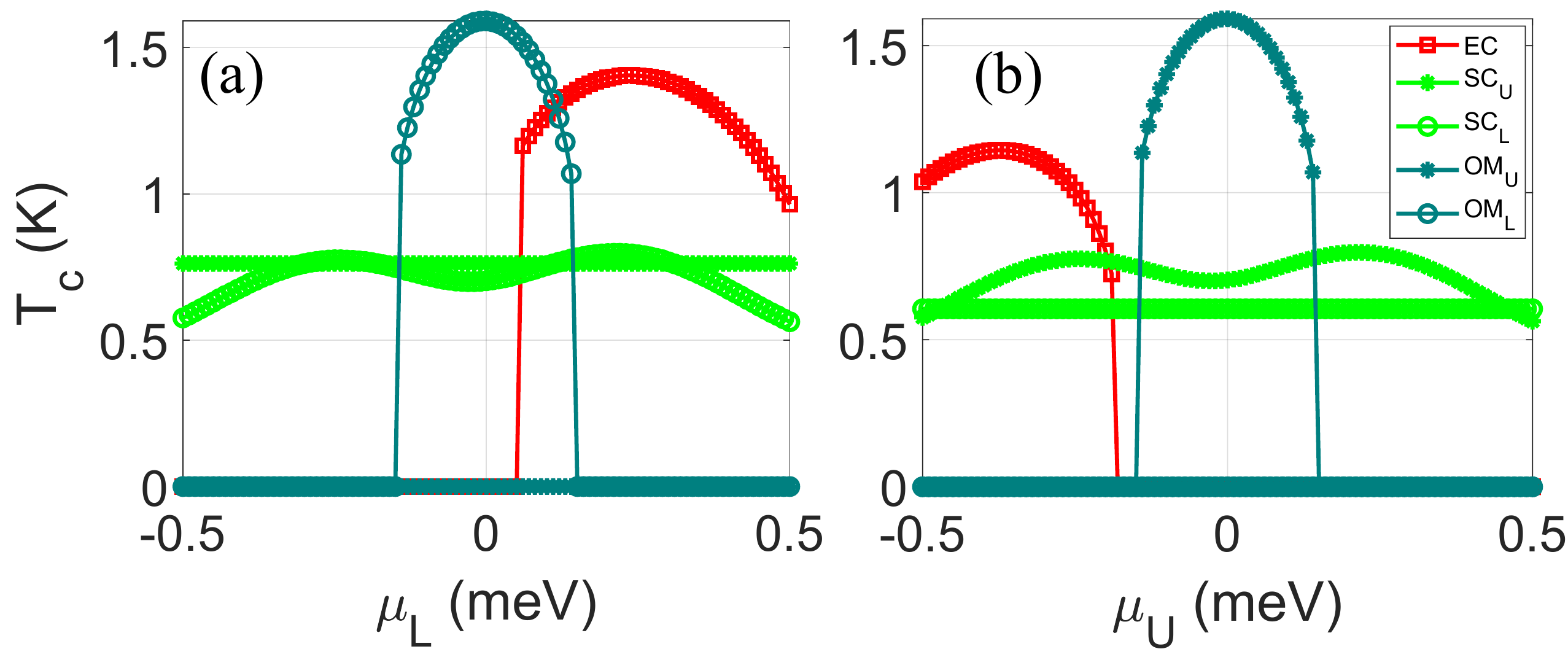}
  \caption{(a) $T_c$ as a function of $\mu_L$
    for $\mu_U=-0.3$~meV, the value used for Fig.~1(c) in the main text.
    (b) $T_c$ as a function of $\mu_U$
    for $\mu_L=0.46$~meV, the value used for Fig.~1(d) in the main text.
    For both plots: $\theta=1.00^\circ$, $V_{\rm EC}=60~$meV$\cdot$nm$^2$,  $V_{\rm SC}=75~$meV$\cdot$nm$^2$, and $V_{\rm OM}=130~$meV$\cdot$nm$^2$.
    }
  \label{fig:1bcut}
 \end{center}
\end{figure}

Figure~1~(b) in the main text, and Fig.~\ref{fig:pdV80V100} show the phase diagrams obtained using the linearized gap equations with the full TBLG bands, as described above. We see that the phase diagrams obtained using the TBLGs' bands, and realistic values of the interaction strengths exhibit some qualitative differences compared to the phase diagram shown in Fig.~\ref{fig:flatpd} obtained using the ideal flat-band model. Figure~\ref{fig:pdV80V100}~(a) shows that  when $V_{\rm EC}= 80$ meV$\cdot$nm$^2$ the region where the EC dominates is almost two thirds of the total area
of the $(\mu_L,\mu_U)$ plane considered, area that corresponds to filling factors ranging from -2.5 to +2.5 for the moir\'e supercell.
When $V_{\rm EC}=100~$meV$\cdot$nm$^2$, the EC phase completely prevails over the other two phases.

Figure~\ref{fig:1bcut} shows the evolution of $T_c$ for the different phases as a function of doping in one TBLG while the doping in the other TBLG is kept fixed.
We see that the SC phase has a finite $T_c$ for all the doping considered, whereas for the EC and OM phase there are range of dopings for which $T_c$ vanishes.


\section{V. Study of phase transitions using the non-linear gap equations}

To investigate the possibility that different order might coexist at low temperature it is necessary to obtain the solution of the full non-linear gap equations in which all the order parameters are present and treated on the same footing.

The general form of the mean-field
Hamiltonian with all the three phases (EC, SC, OM) allowed is:
\begin{equation}
H=\sum_{k}\psi_\kk^\dagger[h_0(\kk))+h_1(\kk)]\psi_\kk,
\label{eq:mftH}
\end{equation}
where
\begin{equation}
\psi_\kk=(\phi_{{\bf K}+\kk U\uparrow},\phi_{{\bf K}+\kk L\uparrow},\phi^\dagger_{-{\bf K}-\kk U\downarrow},\phi^\dagger_{-{\bf K}-\kk L\downarrow})^T,
\end{equation}
with
\begin{eqnarray}
\phi_\kk&=&(c_{\kk+\bb_1 bA},c_{\kk+\bb_1 bB},c_{\kk+\bb_1 tA},c_{\kk+\bb_1 tB},\nonumber\\
&&c_{\kk+\bb_2 bA},c_{\kk+\bb_2 bB},c_{\kk+\bb_2 tA},c_{\kk+\bb_2 tB},\nonumber\\
&&c_{\kk+\bb_3 bA},c_{\kk+\bb_3 bB},c_{\kk+\bb_3 tA},c_{\kk+\bb_3 tB},...),
\end{eqnarray}
and
\begin{eqnarray}
\phi_{-\kk}&=&(c_{-\kk-\bb_1 bA},c_{-\kk-\bb_1 bB},c_{-\kk-\bb_1 tA},c_{-\kk-\bb_1 tB},\nonumber\\
&&c_{-\kk-\bb_2 bA},c_{-\kk-\bb_2 bB},c_{-\kk-\bb_2 tA},c_{-\kk-\bb_2 tB},\nonumber\\
&&c_{-\kk-\bb_3 bA},c_{-\kk-\bb_3 bB},c_{-\kk-\bb_3 tA},c_{-\kk-\bb_3 tB},...), \nonumber\\
\end{eqnarray}
where $\uparrow,\downarrow$ is the spin index, and
$h_0(\kk)$ is the matrix describing the kinetic energy.
To simplify the notation in the remainder we do the following renaming $\kk+{\bf K}\rightarrow \kk$,
$-\kk-{\bf K}\rightarrow -\kk$ in all $c$ and $c^\dagger$ operators. The kernel $h_1(\kk)$ has the form:
\begin{equation}
\small
h_1(\kk)=
\left(
\begin{array}{cc|cc}
\Delta^{U}_{\rm OM}(\kk)&\Delta_{\rm EC}(\kk)&\Delta^U_{\rm SC}(\kk)&0\\
\Delta^{\dagger}_{\rm EC}(\kk)&\Delta^{L}_{\rm OM}(\kk)&0&\Delta^{L}_{\rm SC}(\kk)\\
\hline
\Delta^{U\dagger}_{\rm SC}(\kk)&0&-\Delta^{U}_{\rm OM}(\kk)&-\Delta_{\rm EC}(\kk)\\
0&\Delta^{L\dagger}_{\rm SC}(\kk)&-\Delta^{\dagger}_{\rm EC}(\kk)&-\Delta^{L}_{\rm OM}(\kk)\
\label{eq.h1}
\end{array}
\right).
\end{equation}
In Eq.~\ceq{eq.h1}
$\Delta_{\rm EC}$ is the order parameter for the EC phase,
$\Delta_{\rm OM}^U$ ($\Delta_{\rm OM}^L$)  is the order parameter for the OM phase  in the upper (lower) TBLG,
and $\Delta_{\rm SC}^U$ ($\Delta_{\rm SC}^L$)  is the order parameter for the SC phase  in the upper (lower) TBLG.

We assume the EC order parameter to be periodic on the moir\'e lattice with the Fourier expansion
\begin{equation}
\Delta^{\rm EC}_{\rr l\sigma l'\sigma'}=\sum_b\Delta^{\rm EC}_{\bb l\sigma l'\sigma'}e^{i\bb\cdot\rr},
\end{equation}
where
\begin{equation}
\Delta^{\rm EC}_{\bb l\sigma l'\sigma'}=-\frac{V_{\rm EC}}{\mathcal{A}}\sum_{k_1 b_2}\langle c^\dagger_{L\kk_1+\bb_2l'\sigma'}c_{U\kk_1+\bb_2+\bb l\sigma}\rangle.
\end{equation}

Via a unitary transformation $\mathcal{U}$ into the bands eigenstates,
\begin{equation}
c_{U\kk_1+\bb_1 l\sigma}=\sum_n \mathcal{U}_{Un\kk_1+\bb_1}^{l\sigma}c_{n\kk_1},
\end{equation}
\begin{equation}
c_{L\kk_1+\bb_1 l\sigma}^\dagger=\sum_n \mathcal{U}_{Ln\kk_1+\bb_1}^{l\sigma*}c_{n\kk_1}^\dagger,
\end{equation}
we obtain
\begin{equation}
\Delta^{\rm EC}_{\bb l\sigma l'\sigma'}
=-\frac{V_{\rm EC}}{\mathcal{A}}\sum_{k_1 b_2 m} \mathcal{U}_{Lm\kk_1+\bb_2}^{l'\sigma'*}\mathcal{U}_{Um\kk_1+\bb_2+\bb}^{l\sigma} n_F(E_{m\kk_1}),
\end{equation}
where $E_{m\kk}$ are the eigenvalues of the full Hamiltonian Eq.(\ref{eq:mftH}).
We have checked that this equation is equivalent to the linearized gap Eq.(\ref{eq:LOPec}) in the small order parameter limit.

For the SC order  parameter we have
\begin{eqnarray}
\Delta^{\rm SC}_{\rr l\sigma}
&=&\sum_b\Delta^{\rm SC}_{\bb l\sigma}e^{i\bb\cdot\rr}.
\end{eqnarray}
Assuming the SC order parameter is sub-lattice independent, i.e. $\Delta^{\rm SC}_{\bb lA}=\Delta^{\rm SC}_{\bb lB}\equiv\Delta^{\rm SC}_{\bb l}$ we obtain
\begin{eqnarray}
\Delta^{\rm SC}_{\bb l}
&=&-\frac{V_{\rm SC}}{2\mathcal{A}}\sum_{k_1 b_1\sigma}\langle c_{-(\kk_1+\bb_1)l\sigma\downarrow}c_{\kk_1+\bb_1+\bb l\sigma\uparrow}\rangle.
\end{eqnarray}

By going into the basis that diagonalizes the full Hamiltonian we have
\begin{equation}
c_{\kk_1+\bb_1+\bb l\sigma\uparrow}=\sum_n \mathcal{U}_{n\kk_1+\bb_1+\bb\uparrow}^{l\sigma}c_{n\kk_1},
\end{equation}
\begin{equation}
c_{-(\kk_1+\bb_1)l\sigma\downarrow}=\sum_n \mathcal{U}_{n-(\kk_1+\bb_1)\downarrow}^{l\sigma*}c_{n\kk_1}^\dagger,
\end{equation}
and
\begin{equation}
\Delta^{\rm SC}_{\bb l}
=-\frac{V_{\rm SC}}{2\mathcal{A}}\sum_{k_1 b_1 m\sigma} \mathcal{U}_{m-(\kk_1+\bb_1)\downarrow}^{l\sigma*}\mathcal{U}_{m\kk_1+\bb_1+\bb\uparrow}^{l\sigma} n_F(E_{m\kk_1}).
\end{equation}

For the OM order parameter we have:
\begin{equation}
\Delta^{\rm OM}_{\bb l}=\frac{V_{\rm OM}}{\mathcal{A}}\sum_{k_1 b_1\sigma}\langle c^\dagger_{\kk_1+\bb_1 l\sigma}\hat{\sigma}_z c_{\kk_1+\bb_1+\bb l\sigma}\rangle_{\overline{0}},
\end{equation}
where $\langle\rangle_{\overline{0}}$ means the expectation value excluding the contribution from the non-interacting ground states.
Going into the basis that diagonalizes the full Hamiltonian we obtain:
\begin{eqnarray}
\Delta^{\rm OM}_{\bb l}
&=&\frac{V_{\rm OM}}{\mathcal{A}}\sum_{k_1 b_1 m}[(\mathcal{U}_{m\kk+\bb_1lA}^* \mathcal{U}_{m\kk+\bb_1+\bb lA}\nonumber\\
&&-\mathcal{U}_{m\kk+\bb_1lB}^* \mathcal{U}_{m\kk+\bb_1+\bb lB})n_F(E_{m\kk})\nonumber\\
&&-(\mathcal{U}_{m\kk+\bb_1lA}^{0*} \mathcal{U}^0_{m\kk+\bb_1+\bb lA}\nonumber\\
&&-\mathcal{U}_{m\kk+\bb_1lB}^{0*} \mathcal{U}^0_{m\kk+\bb_1+\bb lB})n_F(\xi^0_{m\kk})],
\end{eqnarray}
where $\xi^0$ are the eigenvalues of the  non-interacting Hamiltonian matrix $h_0$ and
$\mathcal{U}^0$ is the unitary transformation that diagonalizes $h_0$.

To study the phase transitions along
the paths used for Fig.1(c)-(e) in the main text, for each point on such a path, we solve iteratively the non-linear gap equations starting from 30  different, randomly generated, initial states
including states for which all the three order parameters are simultaneously non-zero, and states in which only two, or one, order parameters are not zero.
In order to guarantee that the calculation is fully unconstrained, for each point on the path, we use different seeds to generate the random initial conditions.
We stop the iterative process
when, for each component of the order parameters, the difference of the values between two successive iterations is smaller than $5\times 10^{-4}$~meV.
Among all the solutions obtained from the initial random states, the ground state is identified as the solution for which the total energy per unit cell close to zero temperature
\begin{eqnarray}
E_{tot}&=&\frac{1}{N}\sum_{n\kk}{E_{n\kk}}n_F(E_{nk})+\sum_{\bb l\sigma l'\sigma' s}\frac{|\Delta^{\rm EC}_{\bb l\sigma l'\sigma'}|^2 \mathcal{A}_0}{V_{\rm EC}}\nonumber\\
&&+\sum_{T \bb l\sigma}\frac{|\Delta^{\rm SC}_{\bb lT}|^2 \mathcal{A}_0}{V_{\rm SC}}+\sum_{T \bb ls}\frac{|\Delta^{\rm OM}_{\bb lT}|^2 \mathcal{A}_0}{2V_{\rm OM}},
\end{eqnarray}
is the lowest ($T=(U,L)$ and $s$ is the spin index).

In solving the non-linear gap equation we mesh the Brillouin zone to $2\times(30\times30)$, which is a compromise between speed and precision, with one half of those k-points centering at the origin, and the other half avoiding the high symmetric points in a way similar to the Monkhorst-Pack k-point set~\cite{Pack1976S}. We have verified that further increasing the sample size or the number of k-meshes only leads to minor changes in the magnitudes of the order parameters.

\section{VI. Nearly gapped feature of the exciton condensate state}

\begin{figure}[!htbp]
  \includegraphics[width=\columnwidth]{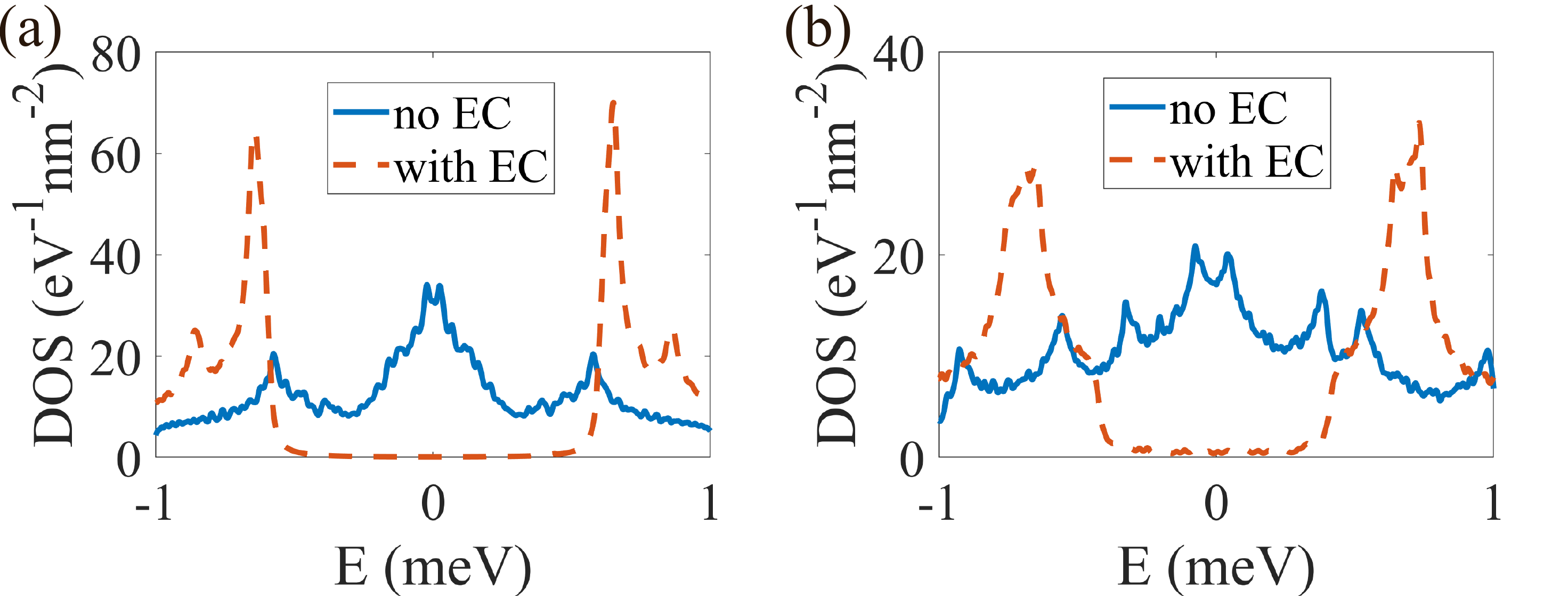}
  \caption{The density of states without and with exciton condensates. Here the parameters are (a)$\theta=1.05\degree$, (b) $\theta=1.00\degree$. In both plots $\mu=0.30~$meV.
  \label{FigS:DOS}}
\end{figure}

From the density of states shown in Fig.~\ref{FigS:DOS}, we can see that although the exciton condensate is a gapless state according to the band structure shown in the main text, the density of states close to the Fermi energy is reduced by more than an order of magnitude in comparison to the normal state. Therefore, in an experiment probing the density of states the transition from normal state to the exciton condensate state would show up as opening of a gap-like feature in the quasiparticle spectrum.

\begin{figure}[!htbp]
\begin{center}
\includegraphics[width=\columnwidth]{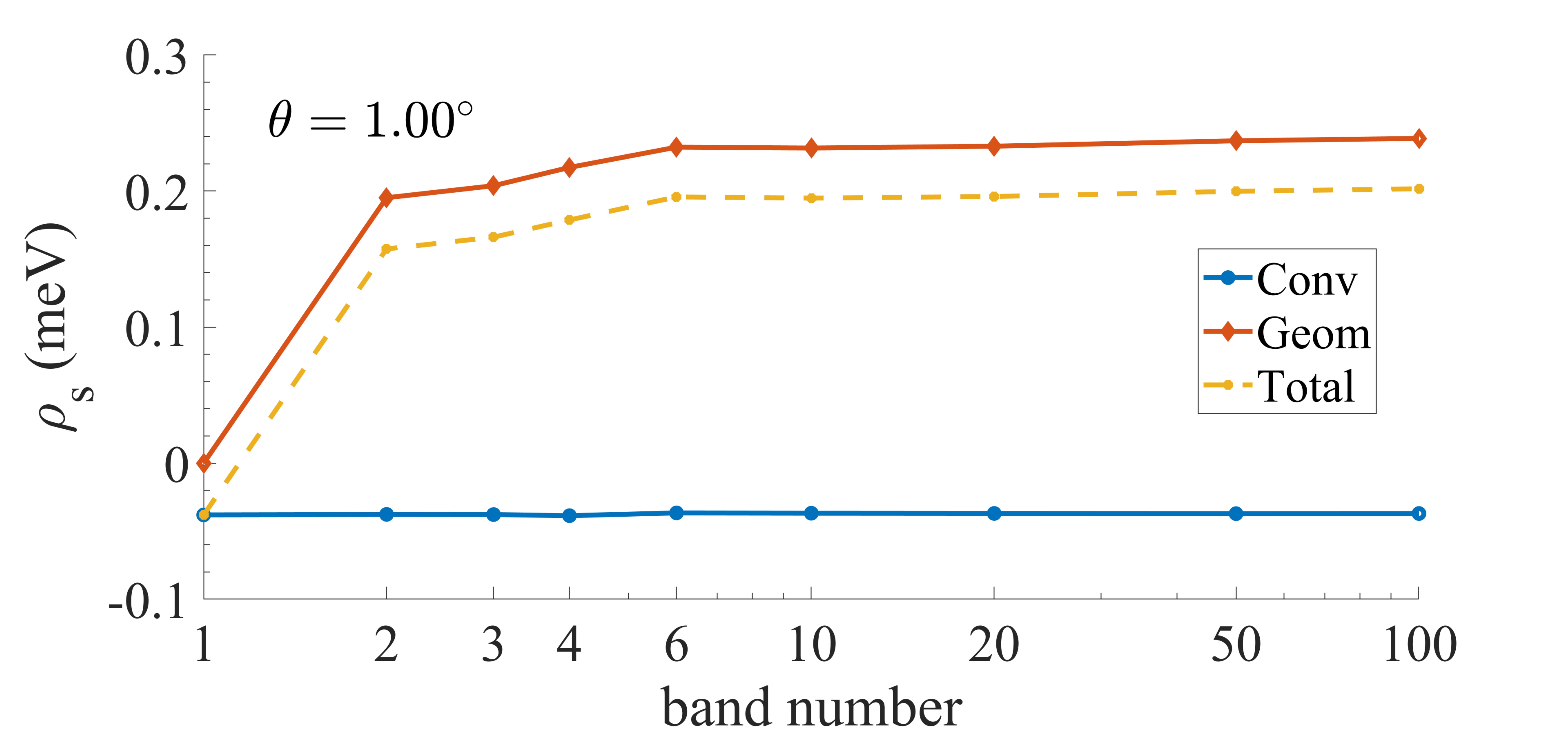}
\caption
{
Here the parameters are $\theta=1.00\degree$, and $\mu=0.30$~meV.
}
\label{FigS:rhosnb}
\end{center}
\end{figure}

\section{VII. Dependence of superfluid density on the number of bands included in the calculation}

Because the geometric part in the superfluid density is purely a multiband effect its value can strongly depend on the number of bands included in the calculation. In Fig.~\ref{FigS:rhosnb}, one can see that for $\theta=1.00\degree$ and $\mu=0.30$meV the conventional part of the superfluid density is almost independent of the number of bands included, in contrast to the geometric part. To get accurate results for the geometric part approximately 10 bands need to be included. We have used 10 bands in all the superfluid density calculations reported in the main text.

\end{document}